\documentclass[aps,pra,twocolumn,showpacs,superscriptaddress]{revtex4}
\usepackage{graphicx}
\usepackage{amsmath}
\usepackage{natbib}
\usepackage{bm}
\usepackage{longtable}
\newcommand{\macro}[1]{\texttt{\textbackslash#1}}
\newcommand{\m}[1]{\macro{#1}}

\begin{document}
\title{Hybrid Quantum Computation}

\author{Arun Sehrawat}
\email[email: ]{arun02@nus.edu.sg}
\affiliation{Centre for Quantum Technologies, National University of Singapore, 3 Science Drive 2, 117543, Singapore}
\affiliation{Department of Physics, National University of Singapore, 2 Science Drive 3, 117542, Singapore}
\author{Daniel Zemann}
\affiliation{Institut f\"{u}r Quantenoptik und Quanteninformation, Technikerstra$\beta$e 21a, A-6020 Innsbruck, Austria}
\altaffiliation[Current address: ]
{Institut f\"{u}r Ionenphysik und Angewandte Physik, Technikerstrasse 25/3, A-6020 Innsbruck, Austria}
\author{Berthold-Georg Englert}
\affiliation{Centre for Quantum Technologies, National University of Singapore, 3 Science Drive 2, 117543, Singapore}
\affiliation{Department of Physics, National University of Singapore, 2 Science Drive 3, 117542, Singapore}

\date{\today}
\pagestyle {plain} %it changes the style from the current page on throughout the remainder of your document. plain: Just a plain page number

%%%%%%%%%%%%%%%%%%%%%%%%%%%%%%%%%%%%%%%%%%%%%%%%%%%%%%%%%%%%%%%%%%%%%%%%%%%%%%%%%%%%%%%%%%%%%%%%%%%%%%%%%%%%%%%%%%%%%%%%%%%%%%%%%%%%

%%%%%%%%%%%%%%%%%%%%%%%%%%%%%%%%%%%%%%%%%%%                                                %%%%%%%%%%%%%%%%%%%%%%%%%%%%%%%%%%%%%%%%%
%%%%%%%%%%%%%%%%%%%%%%%%%%%%%%%%%%%%%%%%%%%                 ABSTRACT                      %%%%%%%%%%%%%%%%%%%%%%%%%%%%%%%%%%%%%%%%%%
%%%%%%%%%%%%%%%%%%%%%%%%%%%%%%%%%%%%%%%%%%%                                                %%%%%%%%%%%%%%%%%%%%%%%%%%%%%%%%%%%%%%%%%

%%%%%%%%%%%%%%%%%%%%%%%%%%%%%%%%%%%%%%%%%%%%%%%%%%%%%%%%%%%%%%%%%%%%%%%%%%%%%%%%%%%%%%%%%%%%%%%%%%%%%%%%%%%%%%%%%%%%%%%%%%%%%%%%%%%%

\begin{abstract}
We present a hybrid model of the unitary-evolution-based quantum computation model and the measurement-based quantum computation model. In the hybrid model part of a quantum circuit is simulated by unitary evolution and the rest by measurements on star graph states, thereby combining the advantages of the two standard quantum computation models. In the hybrid model, a complicated unitary gate under simulation is decomposed in terms of a sequence of single-qubit operations, the controlled-$Z$ gates, and multi-qubit rotations around the $z$-axis. Every single-qubit- and the controlled-$Z$ gate are realized by a respective unitary evolution, and every multi-qubit rotation is executed by a single measurement on a required star graph state. The classical information processing in our model only needs an information flow vector and propagation matrices. We provide the implementation of multi-control gates in the hybrid model. They are very useful for implementing Grover's search algorithm, which is studied as an illustrating example.
\end{abstract}


\maketitle

%%%%%%%%%%%%%%%%%%%%%%%%%%%%%%%%%%%%%%%%%%%%%%%%%%%%%%%%%%%%%%%%%%%%%%%%%%%%%%%%%%%%%%%%%%%%%%%%%%%%%%%%%%%%%%%%%%%%%%%%%%%%%%%%%%%%

%%%%%%%%%%%%%%%%%%%%%%%%%%%%%%%%%%%%%%%%%%%                                                %%%%%%%%%%%%%%%%%%%%%%%%%%%%%%%%%%%%%%%%%
%%%%%%%%%%%%%%%%%%%%%%%%%%%%%%%%%%%%%%%%%%%                 INTRODUCTION                   %%%%%%%%%%%%%%%%%%%%%%%%%%%%%%%%%%%%%%%%%%
%%%%%%%%%%%%%%%%%%%%%%%%%%%%%%%%%%%%%%%%%%%                                                %%%%%%%%%%%%%%%%%%%%%%%%%%%%%%%%%%%%%%%%%

%%%%%%%%%%%%%%%%%%%%%%%%%%%%%%%%%%%%%%%%%%%%%%%%%%%%%%%%%%%%%%%%%%%%%%%%%%%%%%%%%%%%%%%%%%%%%%%%%%%%%%%%%%%%%%%%%%%%%%%%%%%%%%%%%%%%

\section{\label{sec:intro}Introduction}

%%%%%%%%%%%%%%%%%%%%%%%%%%%%%%%%%%%%%                What is a computer?                   %%%%%%%%%%%%%%%%%%%%%%%%%%%%%%%%%% 
A computer is a machine which processes data according to a set of instructions. Every computer is a composition of hardware on which information is processed and software by which information is processed. Hardware is the physical part of a computer, while software is a collection of computer programs (algorithms) designed to perform a required task. A quantum computer (QC) emerges when the computation is executed under the framework of quantum mechanics. There are several models for quantum computation. 

%%%%%%%%%%%%%%%%%%%%%%%%%%%%%%%%%%%%%%%%%            The quantum circuit model                   %%%%%%%%%%%%%%%%%%%%%%%%%%%%%
The first conceptual model of a QC, the quantum Turing machine (QTM), was given by Deutsch \cite{Deutsch85}. It is rather an abstract model and useful for investigating the ``computational complexity." The QTM is a quantum version of its classical analogue which allows the superposition of different computational paths. Later on, a rather practical model of a QC---the unitary-evolution-based quantum computation model (UQCM)---was established \cite{Deutsch89, Barenco95, DiVincenzo}. The UQCM is a quantum edition of ``the reversible classical circuit model." In the step from classical to quantum, the bits are replaced by the qubits, and the logic gates are replaced by the quantum gates (coherent unitary evolution). Unlike the bits, the qubits can exist in a superposition of different computational states. Unlike the logic gates, the quantum gates are able to create and destroy a superposition as well as an entanglement. 

The computation in UQCM is run by a sequence of unitary gates and represented by its circuit diagram, where the connecting wires stand for the logical qubits or bits which carry the information, and the information is processed by the sequence of quantum gates. In the end, the result of the computation is read out by the projective measurements on the qubits. In the well-known textbook by Nielsen and Chuang \cite{NielsenChuang07}, many popular algorithms such as Deutsch's algorithm \cite{Deutsch92}, Grover's search algorithm \cite{Grover97}, and Shor's factoring algorithm \cite{Shor95} are narrated in terms of the UQCM. 
 
%%%%%%%%%%%%%%%%%%%%%%%%%%%%%%%%%%             The one-way quantum computation               %%%%%%%%%%%%%%%%%%%%%%%%%%%%%

The measurement-based quantum computation model (MQCM) is another well recognized model of a QC \cite{Raussendorf01, Raussendorf011}. Here, a multi-qubit entangled state---known as a cluster state \cite{Briegel01}, or more generally a graph state \cite{Briegel02}---is the main ingredient, and it provides all the entanglement beforehand for the subsequent computation. The computation in this model is run by a sequence of single-qubit (adaptive) projective measurements on the graph state. The methods of MQCM enable one to simulate any quantum circuit on a sufficiently large two-dimensional graph state by arranging the spatial pattern of measurement bases for the graph qubits according to the temporal order of quantum gates in the circuit. 

In the MQCM, the measurements on graph qubits are performed in a certain temporal order for the purpose of running the computation deterministically. Furthermore, the measurement outcomes are recorded classically and are used for setting the measurement bases for the subsequent measurements \cite{Raussendorf02}. By contrast, in the UQCM there is no such temporal order of measurements, but an order in which the unitary gates are executed.

%%%%%%%%%%%%%%%%%%%%%%%%%%%%%%%%%%%%%%%%           The Hybrid Quantum Computation             %%%%%%%%%%%%%%%%%%%%%%%%%%%%%

The task of a QC is to ``simulate a quantum circuit.'' Both the UQCM and the MQCM are universal: they can simulate any quantum circuit. Where the UQCM uses the unitary gates, the MQCM uses the measurements for simulating a circuit. In this paper our focus will be on constructing a ``hybrid" model of the UQCM and the MQCM, which we call ``hybrid quantum computation model" (HQCM), that combines elements of both the UQCM and the MQCM with the aim of exploiting the strengths of both models. Of course, the HQCM is universal as well. There are two main objectives of the present investigation. The first objective is to develop a theoretical understanding of the HQCM, where part of a quantum circuit is simulated by unitary gates and the rest by measurements on small graph states. 

The second objective is the investigation of the experimental optimization. Both the UQCM and the MQCM possess their own advantages along with some similarities. For example, where the implementation of an arbitrary single-qubit operation in the MQCM costs a chain of five qubits graph state \cite{Raussendorf01, Raussendorf011}, it can be implemented rather simply by a unitary evolution in the UQCM. In return, certain multi-qubit gates that are complicated in the UQCM can be realized in ``one shot" in the MQCM.

Here is an overview of the paper. In Sec.~\ref{sec:MQCM} we give a brief review of the MQCM, it has two main parts. The first part contains the methodology for the computation in MQCM with an example and the second part is the classical information processing in the MQCM. We use a portion of Sec.~\ref{sec:MQCM} in Sec.~\ref{sec:HQCM}. Section~\ref{sec:HQCM} is reserved for a detailed discussion on the HQCM, it has three main parts. The first part is the methodology for the computation in HQCM and the second part is the classical information processing in the HQCM. The third part is about the implementation of multi-qubit controlled rotations with the HQCM, and this we are going to use in Sec.~\ref{sec:GA}, which is about Grover's search algorithm within the HQCM. Finally, we  conclude our report in Sec.~\ref{sec:Conc} with a summary and outlook. Three appendices deal with technical details.

%%%%%%%%%%%%%%%%%%%%%%%%%%%%%%%%%%%%%%%%%%%%%%%%%%%%%%%%%%%%%%%%%%%%%%%%%%%%%%%%%%%%%%%%%%%%%%%%%%%%%%%%%%%%%%%%%%%%%%%%%%%%%%%%%%%%

%%%%%%%%%%%%%%%%%%%%%%%%%%%%%%%%%%%%%%%%%%%                                                %%%%%%%%%%%%%%%%%%%%%%%%%%%%%%%%%%%%%%%%%
%%%%%%%%%%%%%%%%%%%%%%%%%%%%%%%%%%%%%%%%%%%                 MQCM                            %%%%%%%%%%%%%%%%%%%%%%%%%%%%%%%%%%%%%%%%%%
%%%%%%%%%%%%%%%%%%%%%%%%%%%%%%%%%%%%%%%%%%%                                                %%%%%%%%%%%%%%%%%%%%%%%%%%%%%%%%%%%%%%%%%

%%%%%%%%%%%%%%%%%%%%%%%%%%%%%%%%%%%%%%%%%%%%%%%%%%%%%%%%%%%%%%%%%%%%%%%%%%%%%%%%%%%%%%%%%%%%%%%%%%%%%%%%%%%%%%%%%%%%%%%%%%%%%%%%%%%%

\section{\label{sec:MQCM}A review of the MQCM}

%%%%%%%%%%%%%%%%%%%%%%%%%%%%%%%%%%%%%%%%%%%%%%%%%%%%%%%%%%%%%%%%%%%%%%%%%%%%%%%%%%%%%%%%%%%%%%%%%%%%%%%%%%%%%%%%%%%%%%%%%%%%%%%%%%

%                                                   Methodology for the computation in MQCM

%%%%%%%%%%%%%%%%%%%%%%%%%%%%%%%%%%%%%%%%%%%%%%%%%%%%%%%%%%%%%%%%%%%%%%%%%%%%%%%%%%%%%%%%%%%%%%%%%%%%%%%%%%%%%%%%%%%%%%%%%%%%%%%%%%%

\subsection{\label{sec:Methodology1QC}Methodology for the computation in MQCM}

%%%%%%%%%%%%%%%%%%%%%%%%%%%%%%%%%%%%%                  Cluster state                      %%%%%%%%%%%%%%%%%%%%%%%%%%%%%%%%%%%%%%%%%%

We start this section with a short introduction about the preparation of graph states \cite{Briegel01, Briegel02}, and then we proceed towards the methods for the computation in MQCM \cite{Raussendorf01, Raussendorf011}. We conclude this section with an example. Throughout the text we represent the Pauli vector operator $\vec{\sigma}$ by $\left(X, Y, Z \right)$ and the identity operator by $I$. 

Graph states can be realized in many physical systems by first preparing all the qubits of graph $\mathcal{G}$ in an eigenstate of their respective Pauli operator $X$. In other words, every qubit `$a$' of $\mathcal{G}$ is initialized in the state $\left(\left|0\right\rangle_{a}+(-1)^{\kappa_{a}}\left|1\right\rangle_{a}\right)/\sqrt{2}$, where $ \kappa_{a}\in \left\{0, 1\right\}$. Then, entanglement between each pair of  nearest-neighbor qubits is established by the controlled-$Z$ gate 
\begin{equation}
\textsc{cz}(a,b)= \arrowvert 0\rangle_{a} \langle 0\arrowvert\otimes I^{(b)}+\arrowvert 1\rangle_{a} \langle 1\arrowvert\otimes Z^{(b)}.
\label{cz}
\end{equation}
Here, the indices $a$ and $b$ stand for the qubits at the lattice site `$a$' and its nearest-neighbor lattice site `$b$' of the graph $\mathcal{G}$, respectively. A unitary gate of this kind can be generated by turning on the (controlled) Ising-type nearest-neighbor interaction for an appropriately chosen time period. Experimentally, graph states have been generated by using controlled collisions between cold atoms in optical lattices \cite{Jaksch99} and by using linear optics \cite{linearC1, linearC2, linearC3, linearC4, mbqcGA1, mbqcGA2}.

Mathematically, quantum correlations among the qubits of a graph are specified by correlation operators $K^{(a)}$'s, which are given below. The resultant graph state $\left|\Phi_{\left\{\kappa\right\}} \right\rangle_{\mathcal{G}}$ is an eigenstate of these operators, and it is completely specified by the set of eigenvalue equations
%%%%%%%%%%%%%%%%%%%%%%%%%%%%%%%%%%%%%%%%%%%%%%
\begin{align}
K^{(a)}\left| \Phi_{\left\{\kappa\right\}} \right\rangle_{\mathcal{G}} &=  X^{(a)}\otimes \left(\operatorname*{\bigotimes}_{b\in \mathrm{nbh}(a)}Z^{(b)}\right) \left| \Phi_{\left\{\kappa\right\}} \right\rangle_{\mathcal{G}}  \nonumber \\
& =  (-1)^{\kappa_{a}}\left| \Phi_{\left\{\kappa\right\}} \right\rangle_{\mathcal{G}} \label{eigenvalue}	
\end{align}
%%%%%%%%%%%%%%%%%%%%%%%%%%%%%%%%%%%%%%%%%%%%%%
with the set of eigenvalues $\left\{\kappa\right\}=\left\{\kappa_{a}\in\left\{0, 1\right\}|\:a\in\mathcal{G}\right\}$.
Here, $\mathrm{nbh}(a)$ stands for the set of all nearest-neighbor qubits which are entangled (connected) to the qubit `$a$' by the $\textsc{cz}$ operations. For every qubit `$a$' of the graph state $\left|\Phi_{\left\{\kappa\right\}} \right\rangle_{\mathcal{G}}$, there exists a correlation operator $K^{(a)}$ and an eigenvalue $ \kappa_{a}\in \left\{0, 1\right\}$. The physical meaning of Eq.~(\ref{eigenvalue}) is this: there exists either a correlation ($\kappa_{a} = 0$) or an anti-correlation ($\kappa_{a} = 1$) between the outcome of the measurement on qubit `$a$' in the $X$ eigenbasis and the outcomes of the measurements on all the qubits of $\mathrm{nbh}(a)$ in the $Z$ eigenbasis. These quantum correlations provide the framework for the computation in MQCM. 

%%%%%%%%%%%%%%%%%%%%%%%%%%                  Computation with the MQCM                 %%%%%%%%%%%%%%%%%%%%%%%%%%%%%%%%%

Once the resource graph state is ready, then the logical qubits---holding the input information---are attached to the resource via the same entangling operations given by Eq.~(\ref{cz}). Unlike the quantum error-correction, here, one logical qubit stands for one physical qubit.  Now, the computation is carried out by a sequence of single-qubit (adaptive) projective measurements in a certain direction of the Bloch sphere and in a certain temporal order. The Bloch sphere offers an adequate geometrical description for the direction of single-qubit projective measurement, where the direction of measurement is completely specified by the Bloch vector 
\begin{equation}
\vec{r}(\theta, \varphi) = \left(\sin\theta\cos\varphi,\: \sin\theta\sin\varphi, \:\cos\theta\right), \label{r}
\end{equation}
and the corresponding projector is given by 
\begin{equation}
P_{\vec{r}} = \left(I+ (-1)^{m} \vec{r}\cdot\vec{\sigma}\right)/2. \label{P}
\end{equation}
The measurement outcomes $m = 0$ and $m=1$ mean that the measured qubit is projected onto the states with the kets
\begin{equation}
\left|\uparrow (\theta, \varphi)\right\rangle = \cos(\theta/2)\left|0\right\rangle + e^{i \varphi}\sin(\theta/2)\left|1\right\rangle 
\label{up}
\end{equation}
and 
\begin{equation}
\left|\downarrow(\theta, \varphi)\right\rangle = - \sin(\theta/2)\left|0\right\rangle + e^{i \varphi}\cos(\theta/2)\left|1\right\rangle, \label{down}
\end{equation}
respectively. In other words, the choice of measurement basis is characterized by the direction of measurement $(\theta, \varphi)$ in the Bloch sphere.

There are three kinds of measurements in the MQCM \cite{Raussendorf01, Raussendorf011}. Measurements along the $z$-axis effectively detach the measured (redundant) qubits from the graph state. Measurements along the Bloch vector $\vec{r}_{xy}(\varphi) = (\cos\varphi, \sin\varphi, 0)$---it lies in $x$,$y$ plane of the Bloch sphere---process the information as well as teleport it from one place to another on the graph. The significance of this kind of measurements is revealed by the example given in Appendix \ref{sec:SQR}. Measurements along the Bloch vector $\vec{r}_{zy}(\theta) = (0, \sin\theta, \cos\theta)$---it lies in $z$,$y$ plane of the Bloch sphere---only process the information. This remark will be illustrated by the example given below.

%%%%%%%%%%%%%%%%%%%%%%%%%%%%%%%%%%              Randomness of measurements              %%%%%%%%%%%%%%%%%%%%%%%%%%%%%%%%%%%%%%%%%%%%

The two measurement outcomes $m = 0, 1$ for every qubit of the graph state are equally probable because the reduced density matrix for each qubit is the completely mixed state $I/2$. In the process of getting the desired operations on the logical qubits, one also gets some additional operations. These additional operations are called ``byproduct operators,'' and they belong to the Pauli group. These byproduct operators depend on the random measurement outcomes and the eigenvalues of the graph state $\left|\Phi_{\left\{\kappa\right\}} \right\rangle_{\mathcal{G}}$. The measurement outcome $m\in\left\{0, 1\right\}$ for every graph qubit and the eigenvalues $\left\{\kappa\right\}$ are binary numbers, so one can record them classically in order to take care of the byproduct operators. The classical information processing of this data makes the computation deterministic and helps to set the measurement bases for the subsequent measurements. Section~\ref{sec:CI-MQCM} contains a comprehensive discussion about this matter. Right now we are content with illustrating the procedure of MQCM with an example.

%%%%%%%%%%%%%%%%%%%%%%%%%%%%%%%%%%%%%%     n-qubit GENERALIZED ROTATION       %%%%%%%%%%%%%%%%%%%%%%%%%%%%%%%%%%%%%%%%%%%%%%%%

\begin{figure}[t]% h=here, t=top, p=special page
\centering
\includegraphics[trim = 40mm 60mm 40mm 60mm, clip=true , width=\linewidth]{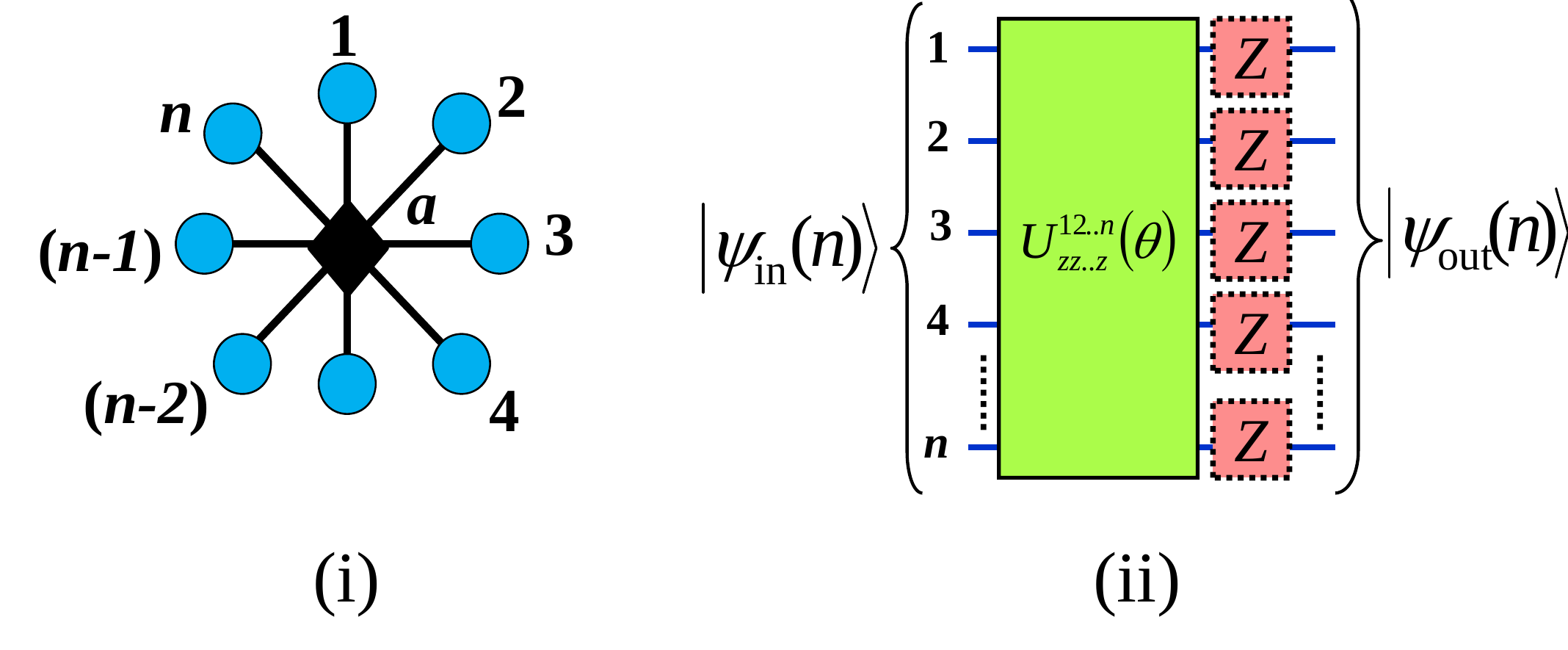}
 %l from the left, b from the bottom, r from the right, and t from the top. Where l, b, r and t are lengths.
\caption{(Color online) (i) This graph is called star graph because of its appearance, and the associated graph state $\arrowvert \phi\rangle_{(1+n)}$ is given by Eq.~(\ref{g2}). The (blue) circles represent the logical qubits which carry the input information; the bonds represent the $\textsc{cz}$ operations given by Eq.~(\ref{cz}), and the ancilla qubit `$a$' is represented by the (black) diamond. Plot~(ii) represents the effect on the input state $\arrowvert \psi_{\mathrm{in}}(n)\rangle$, when the qubit `$a$' of the graph state $\arrowvert\phi\rangle_{(1+n)}$ is measured in an appropriately chosen basis.}\label{fig:1}
\end{figure}

$\textbf{Example:}$ The unitary operation for the $n$-qubit rotation around the $z$-axis is 
\begin{equation}
U^{12...n}_{zz...z}(\theta) = \exp\left(-i\theta Z^{\otimes n}/2\right),
\label{nR}
\end{equation}
where the superscripts $12...n$ symbolize the logical qubits on which this operation will be carried out \cite{Browne06}. One can accomplish this operation by performing a single measurement on a $(1+n)$-qubits star graph state, the associated star graph is shown in Fig.~\ref{fig:1}(i). In Fig.~\ref{fig:1}(i), the input quantum register of $n$ qubits is displayed by the circles and the ancilla qubit `$a$' by the diamond. The input register is in a $n$-qubit input state $\arrowvert \psi_{\mathrm{in}}(n)\rangle$, and the ancilla qubit is prepared in the state $\left(\left|0\right\rangle_{a}+(-1)^{\kappa_{a}}\left|1\right\rangle_{a}\right)/\sqrt{2}$. Then we perform $n$ $\textsc{cz}$ operations, represented by the bonds in the figure and given by Eq.~(\ref{cz}), between the qubit `$a$' and every logical qubit. In principle, all the $\textsc{cz}$ operations can be performed in ``one shot," because they all commute with each other. This series of steps leads us to the resultant star graph state   
\begin{eqnarray}
\arrowvert \phi\rangle_{(1+n)} & = &  \frac{1}{\sqrt{2}} [\arrowvert 0\rangle_{a}\otimes\arrowvert\psi_{\mathrm{in}}(n) \rangle + \nonumber \\
& & \qquad  (-1)^{\kappa_{a}} \arrowvert 1\rangle_{a}\otimes\left(Z^{\otimes n}\arrowvert \psi_{\mathrm{in}}(n)\rangle\right)].
\label{g2}
\end{eqnarray}
The subscript $1+n$ reveals that the final graph state is of one ancilla qubit and $n$ logical qubits.

%%%%%%%%%%%%%%%%%%%%%%%%%%%%%%%%%%%%%%%%%%%%%%%       FIGURE (1)            %%%%%%%%%%%%%%%%%%%%%%%%%%%%%%%%%%%%%%%%%%%%%%%%%

%%%%%%%%%%%%%%%%%%%%%%%%%%%%%%%%%%%%%%%%%%%%%%%%%%%%%%%%%%%%%%%%%%%%%%%%%%%%%%%%%%%%%%%%%%%%%%%%%%%%%%%%%%%%%%%%%%%%%%%%%

A measurement on the ancilla qubit `$a$' in the basis $\left\{\arrowvert \uparrow, \downarrow (\theta, (-1)^{\kappa_{a}}\pi/2)\rangle_{a}\right\}$ transforms the input state of the quantum register into the output state 
\begin{equation}
\arrowvert \psi_{\mathrm{out}}(n)\rangle=\left(Z^{\otimes n}\right)^{m_{a}}U^{12...n}_{zz...z}(\theta)\arrowvert \psi_{\mathrm{in}}(n)\rangle.
\label{out2}
\end{equation}
Here, the direction of measurement lies in $z$,$y$ plane of the Bloch sphere, and $m_{a}\in\left\{0, 1\right\}$ is the measurement outcome. $\left(Z^{\otimes n}\right)^{m_{a}}$ is the byproduct operator, which is represented by the dotted-boxes on all the logical qubits in Fig.~\ref{fig:1}(ii). After the measurement, all bonds (illustrated in Fig.~\ref{fig:1}(i)) are broken and the qubit `$a$' gets projected either onto the state $\arrowvert \uparrow (\theta, (-1)^{\kappa_{a}}\pi/2)\rangle_{a}$
(if $m_{a}= 0$) or onto the state $\arrowvert \downarrow (\theta, (-1)^{\kappa_{a}}\pi/2)\rangle_{a}$
(if $m_{a}= 1$). We shall use this kind of multi-qubit rotations for the HQCM in Sec.~\ref{sec:HQCM}.

In contrast to the example $R_{z}(\varphi)$ given in Appendix~\ref{sec:SQR} where the qubits used for input and output are different, in the case of $U^{12...n}_{zz...z}(\theta)$ the input and output states reside in the same $n$ logical qubits. In other words, here the information gets processed, but does not get transferred from one place to another. 
As a side remark, the resultant byproduct operator $\left(X \right)^{m_{1}}\left(Z\right)^{\kappa_{a}}$ in the case of $R_{z}(\varphi)$ (see Eq.~(\ref{out1})) and the measurement basis $\left\{\arrowvert \uparrow, \downarrow (\theta, (-1)^{\kappa_{a}}\pi/2)\rangle_{a}\right\}$ in the case of $U^{12...n}_{zz...z}(\theta)$ depend on the eigenvalue $\kappa_{a}$.

%%%%%%%%%%%%%%%%%%%%%%%%%%%%%%%
Every quantum gate from the generating set of the Clifford group---the $\textsc{cnot}$ gate
\begin{equation}
\textsc{cnot}(a, b) = \arrowvert 0\rangle_{a} \langle 0\arrowvert\otimes I^{(b)}+\arrowvert 1\rangle_{a} \langle 1\arrowvert\otimes X^{(b)} \label{cnot}
\end{equation}
(the labels $a$ and $b$ are for the control and target qubits, respectively), 
the Hadamard gate 
\begin{equation}
H=(X+Z)/\sqrt{2},\label{H}
\end{equation} 
the $\pi/2$-phase gate 
\begin{equation}
R_{z}(\pi/2)=\exp(-i\pi Z/4)\label{Rz}
\end{equation}---can be executed in a single time step in the MQCM \cite{Raussendorf02}. This holds because every measurement in these cases is performed either in the $X$ eigenbasis or in the $Y$ eigenbasis and is not influenced by the result of any other measurement. Hence all the measurements can be performed simultaneously. A $\textsc{cnot}$ gate can be achieved with a 15-qubit graph state, and both the Hadamard gate and the $\pi/2$-phase gate can be implemented with a chain of five qubits graph state \cite{Raussendorf01, Raussendorf011}. Single-qubit and $\textsc{cnot}$ gates together constitute a universal set of gates, and they are realizable in the MQCM. In this sense, the MQCM is also universal like the UQCM. 

In order to simulate a complex unitary gate in the MQCM, it is customary to first decompose it efficiently into a sequence of elementary gates from the universal gate set. Then, the temporal order of gates is transformed into the spatial pattern of measurement bases for the graph qubits. Afterwards, the measurements are performed in the required order. 

Up to now, we were dealing with the individual gate simulations only, where we need not to worry about the byproduct operators. But in the next section, our focus shall be on the simulation of a sequence of gates, where the study of classical information processing and the temporal order of measurements become necessary. Classical information processing is needed for taking care of the byproduct operators. A comprehensive discussion about it is provided in the following section.

%%%%%%%%%%%%%%%%%%%%%%%%%%%%%%%%%%%%%%%%%%%%%%%%%%%%%%%%%%%%%%%%%%%%%%%%%%%%%%%%%%%%%%%%%%%%%%%%%%%%%%%%%%%%%%%%%%%%%%%%%%%%

%                                       CLASSICAL INFORMATION PROCESSING IN the MQCM

%%%%%%%%%%%%%%%%%%%%%%%%%%%%%%%%%%%%%%%%%%%%%%%%%%%%%%%%%%%%%%%%%%%%%%%%%%%%%%%%%%%%%%%%%%%%%%%%%%%%%%%%%%%%%%%%%%%%%%%%%%%%

\subsection{\label{sec:CI-MQCM}Classical information processing in the MQCM}
This section serves as a summary of the results which were discussed in Ref.~\cite{Raussendorf02}. When a sequence of gates is simulated in the MQCM, the byproduct operator which originates from the implementation of gates ``passes through'' the sequence. The propagation of the byproduct operator either transforms the next gates in the sequence or the byproduct operator in itself gets transformed. The first part of this section is about propagation relations for some elementary gates, and in the second part we shall define an information flow vector. The third part which concludes this section is reserved for the propagation matrices for some elementary gate based on their propagation relations. We shall use a portion of this section for the HQCM in Sec.~\ref{sec:HQCM}. 

The structure of the byproduct operator on the logical qubit $j\in\left\{1, ..., n\right\}$ is  $(X^{(j)})^{x_{j}}(Z^{(j)})^{z_{j}}$, where $x_{j}$ and $z_{j}$ are non-negative integers. Both $x_{j}$ and $z_{j}$ depend on the outcomes of measured qubits and the eigenvalues $\left\{\kappa\right\}$ \cite{Raussendorf02}. Their dependence on $\left\{\kappa\right\}$ is in our control. For example, the $\left\{\kappa\right\}$ dependent parts disappear from the calculation by preparing a graph state with $\kappa=0$ for all the graph qubits. But we cannot control the dependence of the byproduct operators on the measurement outcomes which are intrinsically random.  

In Ref.~\cite{Raussendorf02}, the authors took $x_{j},\;z_{j}\in\left\{0, 1\right\}$, but we find it simpler to take both $x_{j}$ and $z_{j}$ as non-negative integers. This is permissible because in $(X^{(j)})^{x_{j}}$ and $(Z^{(j)})^{z_{j}}$ only the modulo-2 values of $x_{j}$ and $z_{j}$ matter. Throughout the paper, we reserve the sign `+' for the ordinary addition and the sign `$\oplus$' for the modulo-2 addition. 

In principle, we can correct the byproduct operators---step by step---after completing each gate of the sequence under simulation. But it is more convenient to choose not to correct them and let them pass through the gates, and just keep track of the measurement outcomes in a systematic way using simple classical information processing. At the end of the computation, either we set the measurement bases for the final readout depending on the history of outcomes or we just perform the final measurements in the computational basis and interpret the result with the help of the record of measurement outcomes. 

The propagation of the byproduct operator through a gate is given by the propagation relation. The Euler decomposition of an arbitrary single-qubit rotation $R$ is 
\begin{equation}
R(\alpha,\beta,\gamma)=R_{z}(\gamma)R_{x}(\beta)R_{z}(\alpha),\label{Ro}
\end{equation}
and the propagation relation for $R(\alpha,\beta,\gamma)$ is
%%%%%%%%%%%%%%%%%%%%%%%%%%%%%%%%%%%%%%%%%%%%%%%%%%%%
\begin{eqnarray}
\lefteqn{R(\alpha,\beta,\gamma)\:(X)^{x}(Z)^{z} = } \nonumber \\ 
&& (X)^{x}(Z)^{z}\:\tilde{R}\left((-1)^{x}\alpha,(-1)^{z}\beta,(-1)^{x}\gamma\right), \label{proR}
\end{eqnarray}
%%%%%%%%%%%%%%%%%%%%%%%%%%%%%%%%%%%%%%%%%%%%%%%%%%%%
An arbitrary single-qubit rotation $R(\alpha,\beta,\gamma)$ gets transformed to $\tilde{R}\left((-1)^{x}\alpha,(-1)^{z}\beta,(-1)^{x}\gamma\right)$, but the byproduct operator stays as it is. We can take Eq.~(\ref{proR}) as an illustration of the importance of ``the temporal order of the measurements.'' This is because, when $R(\alpha,\beta,\gamma)$ is a part of a circuit, the superscripts $x$ and $z$ are functions of the earlier measurement outcomes. In order to determine the right sign for the measurement angles $\alpha$, $\beta$, and $\gamma$, we have to wait until the necessary measurements are completed \cite{Raussendorf01}. 

Equation~(\ref{proR}) also justifies the following points. The measurement directions for these qubits lie in $x$,$y$ plane of the Bloch sphere, $\vec{r} = (\cos\varphi, \sin\varphi, 0)$ with $\varphi\notin\left\{0, \pm\frac{\pi}{2}\right\}$, their measurement bases depend on the results of previous measurements (see Appendix~\ref{sec:SQR} and Ref.~\cite{Raussendorf01, Raussendorf011}). When $\varphi\in\left\{0, \pm\frac{\pi}{2}\right\}$, then the directions for $+\varphi$ and $-\varphi$ coincide and do not get influenced by the outcomes of other measurements. Measurements of this kind are either in the $X$ ($\varphi=0$) or the $Y$ ($\varphi=\pm\frac{\pi}{2}$) eigenbasis. The gates from the generating set of the Clifford group (given by Eqs.~(\ref{cnot})--(\ref{Rz})) are realized by such measurements. 

%%%%%%%%%%%%%%%%%%%%%%%%%%%%%%%%%%%%%%%%%%%%%%%%%%%%%%%%%%%%%%%%%%%%%%%%%%%%%%%%%%%%%%%%%%%%%%%%%%%%%%%
The propagation relation for the gate $\textsc{cnot}(a, b)$ is
\begin{equation}
\textsc{cnot}(a, b)\textbf{\textit{U}}^{\textsc{cnot}}_{B} =  \tilde{\textbf{\textit{U}}}^{\textsc{cnot}}_{B}\textsc{cnot}(a, b),\label{proCNOT}
\end{equation}
where
\begin{equation}
\textbf{\textit{U}}^{\textsc{cnot}}_{B} = 
(X^{(a)})^{x_{a}}(Z^{(a)})^{z_{a}}(X^{(b)})^{x_{b}}(Z^{(b)})^{z_{b}}, \label{Ucnot}
\end{equation}
and 
\begin{equation}
\tilde{\textbf{\textit{U}}}^{\textsc{cnot}}_{B} = 
(X^{(a)})^{x_{a}}(Z^{(a)})^{z_{a}+ z_{b}}(X^{(b)})^{x_{a}+ x_{b}}(Z^{(b)})^{z_{b}}.\label{Udcnot}
\end{equation}
%%%%%%%%%%%%%%%%%%%%%%%%%%%%%%%%%%%%%%%%%%%%%%%%%%%%%
In case of the $\textsc{cnot}(a, b)$ gate, Eq.~(\ref{proCNOT}), the gate stays as it is, but the byproduct operator $\textbf{\textit{U}}^{\textsc{cnot}}_{B}$ gets transformed to $\tilde{\textbf{\textit{U}}}^{\textsc{cnot}}_{B}$. This is also the case for the other two gates from the generating set of the Clifford group. The propagation relation for the Hadamard gate $H$ is 
\begin{equation}
H(X)^{x}(Z)^{z}=(X)^{z}(Z)^{x}H,\label{proH}
\end{equation}
and for the $\pi/2$-phase gate $R_{z}(\pi/2)$ it is (up to a phase factor $\pm i$)
\begin{equation}
R_{z}(\pi/2)(X)^{x}(Z)^{z}=(X)^{x}(Z)^{z+ x}R_{z}(\pi/2).\label{proPhase} 
\end{equation}
The propagation relations (\ref{proCNOT}), (\ref{proH}), and (\ref{proPhase}) can also be understood from the definition of the Clifford group which maps the Pauli group into itself under conjugation.

Now, let us define an information flow vector. At every stage of the computation, the accumulated byproduct operator $\textbf{\textit{U}}_{B}$ upon the logical qubits $1,..., n$ is of the form $\prod^{n}_{j=1}(X^{(j)})^{x_{j}}(Z^{(j)})^{z_{j}}$. After the implementation of a gate, only the values $\left\{x_{j}\right\}$ and $\left\{z_{j}\right\}$ get changed, and the new values determine the measurement bases for the subsequent gates. These values are processed by a classical computer. There is a one-to-one correspondence between the byproduct operator $\textbf{\textit{U}}_{B}$ (ignoring the global phase $\pm1$) and a $2n$-component ``information flow vector'' $\mathcal{I},$ which is given as follows:
\begin{equation}
\textbf{\textit{U}}_{B}  =  \prod^{n}_{j=1}(X^{(j)})^{x_{j}}(Z^{(j)})^{z_{j}}\Longleftrightarrow \mathcal{I}= 
\left(
\begin{array}{cc}
\mathcal{I}_{x}	\\
\mathcal{I}_{z}
\end{array}
\right), \label{I}
\end{equation}
where
\begin{equation}
\mathcal{I}_{x}  = \left(
\begin{array}{cc}
x_{1}	\\
\vdots\\
x_{n}
\end{array}
\right),\quad
\mathcal{I}_{z}=\left(
\begin{array}{cc}
z_{1}	\\
\vdots\\
z_{n}
\end{array}
\right).\label{Iso}
\end{equation}
Here, the multiplication of byproduct operators (up to a phase factor $\pm1$) corresponds to the component-wise addition of information flow vectors. The information flow vector $\mathcal{I}$ keeps track of the sign(s) of the measurement angle(s) for a gate. In accordance with Eq.~(\ref{proR}), the signs of the measurement angles for the operation $R^{(j)}(\alpha,\beta,\gamma)$ on the qubit $j$ are determined by the current value of $x_{j}$ and $z_{j}$ in $\mathcal{I}$. The propagation relations (\ref{proCNOT}), (\ref{proH}), and (\ref{proPhase}) suggest that none of the gates from the generating set of the Clifford group gets altered under the propagation of the byproduct operator. So, the measurement angles for these gates are independent of the values stored in $\mathcal{I}$. 

We can also define a $2n\times2n$ ``propagation matrix'' $\textbf{C}\left(g\right)$ for a gate $g$, representing the transformation in the information flow vector when the corresponding byproduct operator passes through the gate $g$. 
The propagation matrices given below are derived from the propagation relations (\ref{proR}), (\ref{proCNOT}), (\ref{proH}), and (\ref{proPhase}) with the help of the one-to-one correspondence given by Eq.~({\ref{I}}), and the entries in the information flow vectors and the propagation matrices are given only for relevant qubits. For the case of $n$ logical qubits, the propagation matrices for the $R$-, $\textsc{cnot}$-, $H$-, and $R_{z}(\pi/2)$-gate are given in Appendix~\ref{sec:PM}.

The byproduct operator passes through an arbitrary single-qubit rotation $R(\alpha,\beta,\gamma)$ without getting transformed. So the information flow vector stays as it is, 
%%%%%%%%%%%%%%%%%%%%%%%%%%%%%           PROPAGATION MATRIX for R                 %%%%%%%%%%%%%%%%%%%%%%%%%%%%%%%%%% 
\begin{equation}
\left(
\begin{matrix} % smallmatrix for small matrix
x\\
z\\
\end{matrix}
\right)=\underbrace{\left(
\begin{matrix}
1 & 0\\
0 & 1\\
\end{matrix}
\right)}_{\textbf{C}\left(R\right)}
\left(
\begin{matrix}
x	\\
z\\
\end{matrix}
\right).\label{CR}
\end{equation}
%%%%%%%%%%%%%%%%%%%%%%%%%%%%%%%%%%%%%%%%%%%%%%%%%%%%%%%%%%%%   
The information flow vector gets transformed when the associated byproduct operator passes through the $\textsc{cnot}(a, b)$ gate in the following way:
%%%%%%%%%%%%%%%%%%%%%%%%%%%%%%%%%%%%%%%%             PROPAGATION MATRIX for CNOT             %%%%%%%%%%%%%%%%%%%%%%%%%%%%%%%%%%%
\begin{equation}
\left(
\begin{matrix}
x_{a}	\\
x_{a}+x_{b}\\
z_{a}+z_{b}\\
z_{b}\\
\end{matrix}
\right)
=\underbrace{\left(
\begin{matrix}
1 & 0 & 0 & 0	\\
1 & 1 & 0 & 0	\\
0 & 0 & 1 & 1	\\
0 & 0 & 0 & 1	\\
\end{matrix}
\right)}_{\textbf{C}(\textsc{cnot})}
\left(
\begin{matrix}
x_{a}	\\
x_{b}\\
z_{a}\\
z_{b}
\end{matrix}
\right).\label{Ccnot}
\end{equation}
%%%%%%%%%%%%%%%%%%%%%%%%%%%%%%%%%%%%%%%%     PROPAGATION MATRIX for HADAMARD-gate      %%%%%%%%%%%%%%%%%%%%%%%%%%%%%%%%%%%
Under the one-to-one correspondence given by Eq.~({\ref{I}}), the propagation relation (\ref{proH}) for the Hadamard gate $H$ becomes
\begin{equation}\left(
\begin{matrix}
z\\
x\\
\end{matrix}
\right)
=\underbrace{\left(
\begin{matrix}
0 & 1\\
1 & 0\\
\end{matrix}
\right)
}_{\textbf{C}(H)}
\left(
\begin{matrix}
x	\\
z\\
\end{matrix}
\right),\label{CH}
\end{equation}
%%%%%%%%%%%%%%%%%%%%%%%%%%%%%%%%%%%%%%%%   PROPAGATION MATRIX for PHASE-gate           %%%%%%%%%%%%%%%%%%%%%%%%%%%%%%%%%%%
and the propagation relation (\ref{proPhase}) for the $\pi/2$-phase gate $R_{z}(\pi/2)$ becomes
\begin{equation}
\left(
\begin{matrix}
x\\
z+x\\
\end{matrix}
\right)
=\underbrace{
\left(
\begin{matrix}
1 & 0\\
1 & 1\\
\end{matrix}
\right)}_{\textbf{C}(R_{z}(\pi/2))}
\left(
\begin{matrix}
x	\\
z\\
\end{matrix}
\right).\label{Cphase}
\end{equation}
%%%%%%%%%%%%%%%%%%%%%%%%%%%%%%%%%%%%%%%%%%%%%%%%%%%%%%%%%%%%%%%%%%%%%%%%%%%%%%%%%%%%%%%%%%%%%%%%%%%%%%%%%%%%%%%%%%%%%%%%%%%%%%%%%%%
The information flow vector and the propagation matrices provide a simple description, and they are easily handled by a classical computer.

As a side remark, the temporal order of the measurements does not typically follow the temporal order of gates in a circuit which we want to simulate with the MQCM. In the MQCM, there exists an efficient measurement scheme where measurements are performed round by round, and in each round all the measurements are executed at the same time \cite{Raussendorf02}. The information flow vector is updated after every round. After the final round, the result of the computation is interpreted from the $x$-part of the information flow vector $\mathcal{I}_{x}$. An extended discussion of this efficient measurement scheme is provided in Appendix~\ref{sec:EMS}.

%%%%%%%%%%%%%%%%%%%%%%%%%%%%%%%%%%%%%%%%%%%%%%%%%%%%%%%%%%%%%%%%%%%%%%%%%%%%%%%%%%%%%%%%%%%%%%%%%%%%%%%%%%%%%%%%%%%%%%%%%%%%%%%%%%%%

%%%%%%%%%%%%%%%%%%%%%%%%%%%%%%%%%%%%%%%%%%%                                                %%%%%%%%%%%%%%%%%%%%%%%%%%%%%%%%%%%%%%%%%
%%%%%%%%%%%%%%%%%%%%%%%%%%%%%%%%%%%%%%%%%%%                 HQCM                            %%%%%%%%%%%%%%%%%%%%%%%%%%%%%%%%%%%%%%%%%%
%%%%%%%%%%%%%%%%%%%%%%%%%%%%%%%%%%%%%%%%%%%                                                %%%%%%%%%%%%%%%%%%%%%%%%%%%%%%%%%%%%%%%%%

%%%%%%%%%%%%%%%%%%%%%%%%%%%%%%%%%%%%%%%%%%%%%%%%%%%%%%%%%%%%%%%%%%%%%%%%%%%%%%%%%%%%%%%%%%%%%%%%%%%%%%%%%%%%%%%%%%%%%%%%%%%%%%%%%%%%

\section{\label{sec:HQCM}Hybrid Quantum Computation}

%%%%%%%%%%%%%%%%%%%%%%%%%%%%%%%%%%%%%%%%%%%%%%%%%%%%%%%%%%%%%%%%%%%%%%%%%%%%%%%%%%%%%%%%%%%%%%%%%%%%%%%%%%%%%%%%%%%%%%%%%%%%%%%%%%

%                                       methodology for the computation in HQCM

%%%%%%%%%%%%%%%%%%%%%%%%%%%%%%%%%%%%%%%%%%%%%%%%%%%%%%%%%%%%%%%%%%%%%%%%%%%%%%%%%%%%%%%%%%%%%%%%%%%%%%%%%%%%%%%%%%%%%%%%%%%%%%%%%%%

\subsection{\label{sec:MethodologyHQC} Methodology for the computation in HQCM}

%%%%%%%%%%%%%%%%%%%%%%%%%%%%%%%%%%%%%%%%%%%               Cluster state                %%%%%%%%%%%%%%%%%%%%%%%%%%%%%%%%%%%%%%%%%%%%%
In this section, we formulate the methodology of the hybrid quantum computation model (HQCM) by combining the advantages of the UQCM and the MQCM. Before going into the details, let us first focus on what benefits we can get from each of the models in different situations. Here, we consider the preparation of a graph state, the set of elementary gates for the HQCM, and the simulation of a quantum circuit with the HQCM one by one. 

The very first experimental step in the MQCM is the preparation of a resource graph state, whereas in the UQCM no such preparation of a resource is needed. While preparing a graph state, in principle, the initialization of every graph qubit in the $X$ eigenbasis can be completed in one shot. To this end, we have to ``talk'' to every graph qubit simultaneously. Consequently, this requires a lot of experimental resources, and the very many interactions are difficult to control. Likewise, the subsequent two-qubit entangling operations ($\textsc{cz}(a, b)$'s given by Eq.~(\ref{cz})) to create the resource graph state can be performed in one step, because they commute with each other. And, the larger the graph state, the more difficult it is to prepare and control the state and to protect it against decoherence. So, for the HQCM, we choose not to prepare the whole two-dimensional universal graph state at once, but instead prepare small (non-universal) graph states step-by-step as we need them when the computation progresses. Only the star graph states, such as $\arrowvert \phi\rangle_{(1+n)}$ given in Eq.~(\ref{g2}), are required for the HQCM.

%%%%%%%%%%%%%%%%%%%%%%%%%%%%%%%%%%%%%%%              elementary gates           %%%%%%%%%%%%%%%%%%%%%%%%%%%%%%%%%%%%%%%%%%%%%

We choose single-qubit operations, the $\textsc{cz}$ gate and the multi-qubit rotation around the $z$-axis  $U^{12...n}_{zz...z}(\theta)$ for an arbitrary value of $\theta$ given by Eq.~(\ref{nR}) as the elementary gates for HQCM. In analogy to the procedure for the UQCM, first we ``efficiently decompose'' any big unitary gate we are trying to simulate into a sequence of elementary gates in such a way that the number of elementary gates grows polynomially with the number of logical qubits, and then every elementary gate is implemented one after another. Every single-qubit operation and the $\textsc{cz}$ gate are carried out by the unitary evolution under the formalism of UQCM. The rotation $U^{12...n}_{zz...z}(\theta)$ is implemented by the method given in Sec.~\ref{sec:Methodology1QC} (see the example) under the formalism of MQCM. The motivation behind these choices is explained in the following. 

The implementation of an arbitrary single-qubit rotation in the MQCM costs us at least a chain of five qubits graph state and four measurements \cite{Raussendorf01, Raussendorf011}. But it can be realized quite simply by the unitary evolution of the respective single qubit. Furthermore, the Euler decomposition for an arbitrary single-qubit rotation $R(\alpha,\beta,\gamma)$ given in Eq.~(\ref{Ro}) is not needed. 

The $\textsc{cz}$ operations themselves are part of the experimental setup for constructing the graph states, and for this we have to execute them by the unitary evolution. That is why we consider the $\textsc{cz}$ gate as an elementary gate for the HQCM. Furthermore, it is more economical to implement $\textsc{cnot}(a, b)$ by the unitary evolution $H^{(b)}\textsc{cz}(a, b)H^{(b)}$ instead of first preparing a 15-qubit graph state and then implement it with the MQCM \cite{Raussendorf01, Raussendorf011}. 

Although we already have the universal set of gates (single-qubit and $\textsc{cz}$ gates), we prefer to include $U^{12...n}_{zz...z}(\theta)$ as an elementary gate in the HQCM. This is because of two reasons. The first reason is the optimization. The resource $(1+n)$-qubit graph state $\arrowvert \phi\rangle_{(1+n)}$ (given by Eq.~(\ref{g2})) needed for the implementation of $U^{12...n}_{zz...z}(\theta)$ is relatively easy to create experimentally. It has only one ancilla qubit, and the entanglement can be established in one shot. Furthermore, a single measurement on the ancilla qubit is enough to realize $U^{12...n}_{zz...z}(\theta)$ all together on $n$ logical qubits. While it is also possible to decompose the rotation $U^{12...n}_{zz...z}(\theta)$ in terms of the gates from the universal gate set and implement it under the formalism of UQCM, its implementation there will not be so optimal, and we cannot regard it as a single unit. 

Generally, one is using either unitary evolution (UQCM) or measurements on the graph state (MQCM) in order to simulate a quantum circuit. So, the second reason for including $U^{12...n}_{zz...z}(\theta)$ as an elementary gate in the HQCM is to investigate a model of QC in which a part of a circuit ($U^{12...n}_{zz...z}(\theta)$ rotations) is simulated by the measurements and the rest by the unitary evolution, and to see how the classical information processing plays its role in such a model. The importance of looking at the classical information-processing parts is explained in the following.

%%%%%%%%%%%%%%%%%%%%%%%%%%%%%                 classical information processing               %%%%%%%%%%%%%%%%%%%%%%%%%%%%%%%%%

Now let us consider the simulation of a sequence of gates with the HQCM in which the classical information processing becomes crucial. The classical processing does not come into the picture of UQCM where the measurements are used only for the readout of the final result of computation. In all those schemes where measurements are needed for the computation (for example quantum teleportation \cite{teleport}), the classical information processing in parallel is essential. In the HQCM also, classical information processing is needed, because the rotations $U^{12...n}_{zz...z}(\theta)$ are executed by the measurements. But here the classical information-processing parts are rather simple and straightforward, requiring only the information flow vector and the propagation matrices. A comprehensive discussion of this is given in the following section.
%
%
%%%%%%%%%%%%%%%%%%%%%%%%%%%%%%%%%%%%%%%%%%%%%%%%%%%%%%%%%%%%%%%%%%%%%%%%%%%%%%%%%%%%%%%%%%%%%%%%%%%%%%%%%%%%%%%%%%%%%%%%%%%%
%
%                                       CLASSICAL INFORMATION PROCESSING in the HQCM
%
%%%%%%%%%%%%%%%%%%%%%%%%%%%%%%%%%%%%%%%%%%%%%%%%%%%%%%%%%%%%%%%%%%%%%%%%%%%%%%%%%%%%%%%%%%%%%%%%%%%%%%%%%%%%%%%%%%%%%%%%%%%%
%
\subsection{\label{sec:CI-HQCM}Classical information processing in the HQCM}
In this section, our focus will be on the classical information-processing parts of the HQCM. We only need the information flow vector and the propagation matrices for the elementary gates. We first redefine the information flow vector in the context of HQCM, and then discuss the propagation relations as well as the propagation matrices for the elementary gates. 

At every computation step $\tau$, the form of the byproduct operator is the same, as given in Eqs.~(\ref{I}) and (\ref{byproMBQCM}),
 \begin{equation}
\textit{\textbf{U}}_{B}(\tau)=\prod^{n}_{j=1}(X^{(j)})^{x_{j}(\tau)}(Z^{(j)})^{z_{j}(\tau)}.\label{byproHQCM} 
\end{equation}
So, the form of the related information flow vector $\mathcal{I}(\tau)=\mathcal{I}({x_{j}(\tau), z_{j}(\tau)})$ is also the same as given by the $2n\times1$ column vector in Eqs.~(\ref{I}) and (\ref{Iso}). But in the HQCM, there are some differences in comparison to the efficient measurement scheme of MQCM given in Appendix~\ref{sec:EMS}. In that scheme of MQCM, the index `$t$' of $\mathcal{I}(t)$ stands for `the measurement round.' But in the HQCM, every elementary gate is taken as a single computational step, and the index `$\tau$' of $\mathcal{I}(\tau)$ is the label for them. In the MQCM, $\mathcal{I}(t)$ gets updated after each round, but in the HQCM it is updated after each gate.  

In this scheme of MQCM, the initial value of the information flow vector $\mathcal{I}^{\textsc{mqcm}}_{\textrm{init}}$ is determined by the set of eigenvalues $\left\{\kappa\right\}$ and by some particular gates. But in the HQCM, just before starting the computation all the entries of $\mathcal{I}(0)=\mathcal{I}^{\textsc{hqcm}}_{\textrm{init}}$ are zeros, i.e., both $x_{j}(0)=0$ and $z_{j}(0)=0$ for all $j = 1,2,\cdots,n$. That means that the byproduct operator at $\tau=0$ is the identity operator $I$ on every logical qubit. In fact, the first relevant byproduct operator appears in the computation when the first multi-qubit rotation is implemented, and then the information flow vector gets some nonzero entries. In the MQCM, the information flow vector gets updated from $\mathcal{I}(t-1)$ to $\mathcal{I}(t)$ after the $t$th measurement round. In the HQCM, the information flow vector gets updated from $\mathcal{I}(\tau-1)$ to $\mathcal{I}(\tau)$ after the implementation of $\tau$th gate. $\mathcal{I}(\tau)$ influences the $(\tau+1)$th gate of a quantum circuit under simulation. Similar to the case of the UQCM, the total number of computation steps (the logical depth) is denoted by $\tau_{\textrm{max}}$, which is the total number of elementary gates used for the computation. Furthermore, $\tau_{\textrm{max}}$ is also the total number of steps taken by a classical computer for the classical information processing in the HQCM.

Let us turn to the issue how one can interpret the final result of the computation in the HQCM. In the case of UQCM, every gate of a circuit is executed by their respective unitary evolution, and the final readout measurements are performed in the computational basis. In this case, the output state $\left|\mathrm{out}\right\rangle$ gets projected onto the state $\left|M_{\mathrm{UQCM}}\right\rangle = \otimes^{n}_{j=1}\left|\acute{s}_{j}\right\rangle$ after the final readout measurements, 
\begin{equation}
\left|M_{\mathrm{UQCM}}\right\rangle = \prod^{n}_{j=1}\frac{I^{(j)}+(-1)^{\acute{s}_{j}}Z^{(j)}}{2}\left|\mathrm{out}\right\rangle,
\label{outU} 
\end{equation}
where $\acute{s}_{j}\in\left\{0, 1\right\}$ are the readout measurement outcomes for the logical qubits $j=1, 2, \cdots, n$. 

In the case of HQCM, the final state of the output register will be $\textbf{\textit{U}}_{B}(\tau_{\mathrm{max}})\left|\mathrm{out}\right\rangle$ after performing the last gate of the same circuit. Without loss of generality, like above, we consider the computational basis for the final readout, where $s_{j}\in\left\{0, 1\right\}$ are the readout measurement outcomes for the logical qubits $j=1, 2, \cdots, n$. That means that the output state $\textbf{\textit{U}}_{B}(\tau_{\mathrm{max}})\left|\mathrm{out}\right\rangle$ gets projected onto the state $\left|M_{\mathrm{HQCM}}\right\rangle = \otimes^{n}_{j=1}\left|s_{j}\right\rangle$ after the readout measurements, 
\begin{equation}
\left|M_{\mathrm{HQCM}}\right\rangle = \prod^{n}_{j=1}\frac{I^{(j)}+(-1)^{s_{j}}Z^{(j)}}{2}\textbf{\textit{U}}_{B}(\tau_{\mathrm{max}})\left|\mathrm{out}\right\rangle. \label{outH}
\end{equation}
We can transform Eq.~(\ref{outH}) with the help of Eq.~(\ref{byproHQCM}) into
\begin{equation}
\left|M_{\mathrm{HQCM}}\right\rangle = \textbf{\textit{U}}_{B}(\tau_{\mathrm{max}})\prod^{n}_{j=1}\frac{I^{(j)}+(-1)^{s_{j}+x_{j}(\tau_{\mathrm{max}})}Z^{(j)}}{2}\left|\mathrm{out}\right\rangle. \label{out2H}
\end{equation}

The inference we get by comparing Eq.~(\ref{outU}) and Eq.~(\ref{out2H}) is: the readout measurements on the state $\left|\mathrm{out}\right\rangle$ with the results $\left\{\acute{s}_{j}\right\}$ give the same circuit-output as the readout measurements on the state $\textbf{\textit{U}}_{B}(\tau_{\mathrm{max}})\left|\mathrm{out}\right\rangle$ with the results $\left\{s_{j}\right\}$, and these sets of results are related by
\begin{equation}
\acute{s}_{j} \equiv s_{j} + x_{j}(\tau_{\mathrm{max}}) ~~\mbox{for all}~~ j\in\left\{1, 2, \cdots, n\right\}.
\label{sx}
\end{equation}
That is how one can interpret the final result of the computation with the help of $\mathcal{I}_{x}(\tau_{\textrm{max}})$ in the HQCM.

%%%%%%%%%%%%%%%%%%%%%%%%%%%         Propagation relation for single qubit rotation         %%%%%%%%%%%%%%%%%%%%%%%%%%%%%%%%%%%%%

Let us turn to the propagation relations for the elementary gates. An arbitrary single-qubit rotation around an axis $\vec{r}(\theta,\varphi)$ (defined in Eq.~(\ref{r})) by an angle $\alpha$ is 
\begin{equation}
R_{\vec{r}}(\alpha)=\exp\left(-i\frac{\alpha}{2}\vec{r}.\vec{\sigma}\right). 
\label{R}
\end{equation}
The byproduct operator passes through this gate without any change, but it changes the axis of rotation of the gate from $\vec{r}$ to $\vec{r}\:'$. The propagation relation for $R_{\vec{r}}(\alpha)$ is given by 
%%%%%%%%%%%%%%%%%%%%%%%%%%%%%%%%%%%%%%%%%%%%%%%%%%%%
\begin{equation}
R_{\vec{r}}(\alpha)(X)^{x}(Z)^{z}  =(X)^{x}(Z)^{z}R_{\vec{r}\:'}(\alpha), 
\label{HQCproR}	
\end{equation}
where 
\begin{equation}
\vec{r}\;'= \left((-1)^{z}\sin\theta\cos\varphi, (-1)^{x+z}\sin\theta\sin\varphi, (-1)^{x}\cos\theta\right).
\label{rd}
\end{equation}
%%%%%%%%%%%%%%%%%%%%%%%%%%%%%%%%%%%%%%%%%%%%%%%%%%%
In other words, the angles $\theta$, $\varphi$ that define the axis of rotation $\vec{r}$ get transformed as $\theta\rightarrow \left(x\pi-\theta\right)$ and $\varphi\rightarrow (-1)^{x}\left(z\pi+\varphi\right)$. The byproduct operator passes through $R_{\vec{r}}(\alpha)$ without getting transformed, that means that the propagation matrix $\textbf{C}(R)$ is the same identity matrix as given in Eqs.~(\ref{CR}) and (\ref{CCR}). Every single-qubit unitary operator in SU(2) follows this propagation relation, and it becomes the propagation relation (\ref{proH}) for the Hadamard gate when $\theta=\varphi=\pi/2$ and the propagation relation (\ref{proPhase}) for the $\pi/2$-phase gate when $\theta=0$. So, the Hadamard- and the $\pi/2$-phase-gate remain special cases in the sense that the byproduct operator changes under the propagation but not these gates. They are executed by the unitary evolution like any other single-qubit gate, but for the classical information-processing parts of the HQCM we shall use their propagation matrices given by Eqs.~(\ref{CCH}) and (\ref{CCphase}).
 
The propagation relation for the next elementary gate $\textsc{cz}(a, b)$ (defined in Eq.~(\ref{cz})) is
%%%%%%%%%%%%%%%%%%%%%%%%%%%%%%
\begin{equation}
\textsc{cz}(a, b)\textbf{\textit{U}}^{\textsc{cz}}_{B} =  \tilde{\textbf{\textit{U}}}^{\textsc{cz}}_{B}\textsc{cz}(a, b),\label{HQCproCZ}
\end{equation}
where
\begin{equation}
\textbf{\textit{U}}^{\textsc{cz}}_{B}  = 
(X^{(a)})^{x_{a}}(Z^{(a)})^{z_{a}}(X^{(b)})^{x_{b}}(Z^{(b)})^{z_{b}},\label{Ucz}
\end{equation}
and
\begin{equation}
\tilde{\textbf{\textit{U}}}^{\textsc{cz}}_{B}  = 
(X^{(a)})^{x_{a}}(Z^{(a)})^{z_{a}+ x_{b}}(X^{(b)})^{x_{b}}(Z^{(b)})^{z_{b}+ x_{a}},\label{Udcz}
\end{equation}
%%%%%
%%%%%%%%%%%%%%%%%%%%%%%%%%%%%%%%%%%%%%%%%%%%%%%%%%%%%
Under the one-to-one correspondence given in Eq.~(\ref{I}) the propagation relation (\ref{HQCproCZ}) becomes
%%%%%%%%%%%%%%%%%%%%%%%%%%%%%%%%%%%%%%%%             PROPAGATION MATRIX for CZ             %%%%%%%%%%%%%%%%%%%%%%%%%%%%%%%%%%%
\begin{equation}\left(
\begin{matrix} %smallmatrix
x_{a}	\\
x_{b}\\
z_{a}+x_{b}\\
z_{b}+x_{a}\\
\end{matrix}
\right)
=\underbrace{\left(
\begin{matrix}
1 & 0 & 0 & 0	\\
0 & 1 & 0 & 0	\\
0 & 1 & 1 & 0	\\
1 & 0 & 0 & 1	\\
\end{matrix}
\right)}_{\textbf{C}(\textsc{cz})}
\left(
\begin{matrix}
x_{a}	\\
x_{b}\\
z_{a}\\
z_{b}
\end{matrix}
\right).\label{Ccz}
\end{equation}  
When the control qubit `$a$' and the target qubit `$b$' belong to the set of $n$ logical qubits $(a\neq b)$, then the propagation matrix $\textbf{C}(\textsc{cz}(a, b))$ can be generated by the following relations:
\begin{align}
\left[\textbf{C}_{xx}(\textsc{cz}(a, b))\right]_{kl}&= \left[\textbf{C}_{zz}(\textsc{cz}(a, b))\right]_{kl}=\delta_{kl},\nonumber\\
\left[\textbf{C}_{xz}(\textsc{cz}(a, b))\right]_{kl}&=\delta_{ka}\delta_{lb}+\delta_{kb}\delta_{la},\nonumber\\
\left[\textbf{C}_{zx}(\textsc{cz}(a, b))\right]_{kl}&= 0.\label{CCcz}
\end{align}
Note that Eqs.~(\ref{Ccz}) and (\ref{CCcz}) are different from Eqs.~(\ref{Ccnot}) and (\ref{CCcnot}).
The $\textsc{cz}$ and $\textsc{cnot}$ gates are interconvertible by using the Hadamard gate, i.e., $H^{(b)}\textsc{cz}(a, b)H^{(b)}=\textsc{cnot}(a, b)$ and the same is true for their propagation matrices, i.e., 
\begin{equation}
\textbf{C}(H^{(b)})\textbf{C}(\textsc{cz}(a, b))\textbf{C}(H^{(b)})=\textbf{C}(\textsc{cnot}(a, b)).\label{HczH}
\end{equation}

%%%%%%%%%%%%%%%%%%%%%%%%%%%     Propagation relation for multi-qubit rotation           %%%%%%%%%%%%%%%%%%%%%%%%%%%%%%%%%%%%%%%%

The propagation relation for $U^{12...n}_{zz...z}(\theta)$ is
%%%%%%%%%%%%%%%%%%%%%%%%%%%%%
\begin{equation}
U^{12...n}_{zz...z}(\theta)\textbf{\textit{U}}_{B} = \textbf{\textit{U}}_{B}U^{12...n}_{zz...z}((-1)^{x}\theta),\label{HQCproU}
\end{equation}
where $\textbf{\textit{U}}_{B}$ is the same as given in Eq.~(\ref{I}) and 
\begin{equation}
x  = \sum^{n}_{j=1}x_{j}.\label{x}
\end{equation}
%%
%%%%%%%%%%%%%%%%%%%%%%%%%%%%%%%%%%%%%%%%%%
In this case, the measurement angle $\theta$ gets modified under the propagation, but the byproduct operator stays as it is. So, the propagation matrix $\textbf{C}(U^{12...n}_{zz...z}(\theta))$ will be the identity matrix, which can be defined in the same way as the  $\textbf{C}(R)$ in Eq.~(\ref{CCR}). The inferences we get here are the following: (1) The Hadamard-, the $\pi/2$-phase- and the $\textsc{cz}$-gate remain unchanged under the propagation, while the byproduct operator gets altered. (2) The Single- and multi-qubit rotations (with nontrivial angles) get transformed, while the byproduct operator stays unaltered under the propagation. 

Now we have all the basic tools for the HQCM. We shall first describe some important examples, and then proceed to the implementation of Grover's search algorithm within the HQCM.

%%%%%%%%%%%%%%%%%%%%%%%%%%%%%%%%%%%%%%%%%%%%%%%%%%%%%%%%%%%%%%%%%%%%%%%%%%%%%%%%%%%%%%%%%%%%%%%%%%%%%%%%%%%%%%%%%%%%%%%%%%%%

%                                                     CONTROLLED OPERATIONS

%%%%%%%%%%%%%%%%%%%%%%%%%%%%%%%%%%%%%%%%%%%%%%%%%%%%%%%%%%%%%%%%%%%%%%%%%%%%%%%%%%%%%%%%%%%%%%%%%%%%%%%%%%%%%%%%%%%%%%%%%%%%

\subsection{\label{sec:control}Controlled operations with the HQCM}
In this section we are considering $n$-qubit controlled rotations around the $z$-axis, which are defined by
\begin{widetext}  %change the formatting from two-column to one-column to better accommodate very long equations
\begin{equation}
\Lambda^{1...c}U^{(c+1)...n}_{z...z}(\theta)=\left(I^{\otimes c}-\left|1...1\right\rangle_{1...c}\left\langle 1...1\right|\right)\otimes I^{\otimes (n-c)} 
  +\left|1...1\right\rangle_{1...c}\left\langle 1...1\right|\otimes U^{(c+1)...n}_{z...z}(\theta),\label{ClU}
\end{equation} 
\end{widetext}
where the qubits labeled 1 to $c$ are the control qubits and the qubits labeled $c+1$ to $n$ are the target qubits. Only if every control qubit is in the state $\left|1\right\rangle$, then the $(n-c)$-qubit rotation $U^{(c+1)...n}_{z...z}(\theta)$ operates on the target qubits. We discuss the HQCM implementation for three different values of $c$: $c$ = 1 (single-control), $c$ = 2 (double-control) and $c = 3$ (triple-control). 

%%%%%%%%%%%%%%%%%%%%%%%%%%%%%%%%%%%%%%%%%%%%%%%            CASE (1): one-control         %%%%%%%%%%%%%%%%%%%%%%%%%%%%%%%%%%

\subsubsection{\label{sec:C1} The single-control gate $\Lambda^{1}U^{2...n}_{z...z}(-2\theta)$}

First, we decompose $\Lambda^{1}U^{2...n}_{z...z}(-2\theta)$ in terms of multi-qubit rotations like $U^{12...n}_{zz...z}(\theta)$ given by Eq.~(\ref{nR}). In order to have the logical qubit 1 as the control and the rest as the target qubits, we express the $n$-qubit rotation about the $z$-axis as $$U^{12...n}_{zz...z}(\theta)=\arrowvert 0\rangle_{1}\langle 0\arrowvert \otimes U^{2...n}_{z...z}(\theta)+\arrowvert 1\rangle_{1}\langle 1\arrowvert \otimes U^{2...n}_{z...z}(-\theta).$$ Consequently, we get the required decomposition:
%%%%%%%%%%%%%%%%%%%%%%%%%%%%%%%%%%%              C1-U             %%%%%%%%%%%%%%%%%%%%%%%%%%%%%%%%%%%%%%%%%%%%%%%%%%%%%%%%%%
\begin{equation}
\Lambda^{1}U^{2...n}_{z...z}(-2\theta)  =  U^{2...n}_{z...z}(-\theta) U^{12...n}_{zz...z}(\theta). \label{C1U}
\end{equation}
%\nonumber \\
%&  = & \arrowvert 0\rangle_{1}\langle 0\arrowvert \otimes\textit{\textbf{I}}^{\otimes (n-1)}  +\arrowvert 1\rangle_{1}\langle 1\arrowvert \otimes U^{2...n}_{z...z}(-2\theta).
%%%%%%%%%%%%%%%%%%%%%%%%%%%%%%%%%%%%%%%%%%%%%%%%%%%%%%%%%%%%%%%%%%%%%%%%%%%%%%%%%%%%%%%%%%%%%%%%%%%%%%%%%%%%%%%%%%%%%%%%%%%%
We observe that $U^{12...n}_{zz...z}(\theta)$ is symmetric under permutation of the qubits, so we can take any qubit as the control and the remaining qubits as targets. We consider a `multi-qubit rotation about the $z$-axis' as a single unit, and then $\Lambda^{1}U^{2...n}_{z...z}(-2\theta)$ costs only two units of this kind.

\begin{figure}[t]
\centering
\includegraphics[ trim = 35mm 35mm 42mm 45mm, clip=true , width=0.8\linewidth]{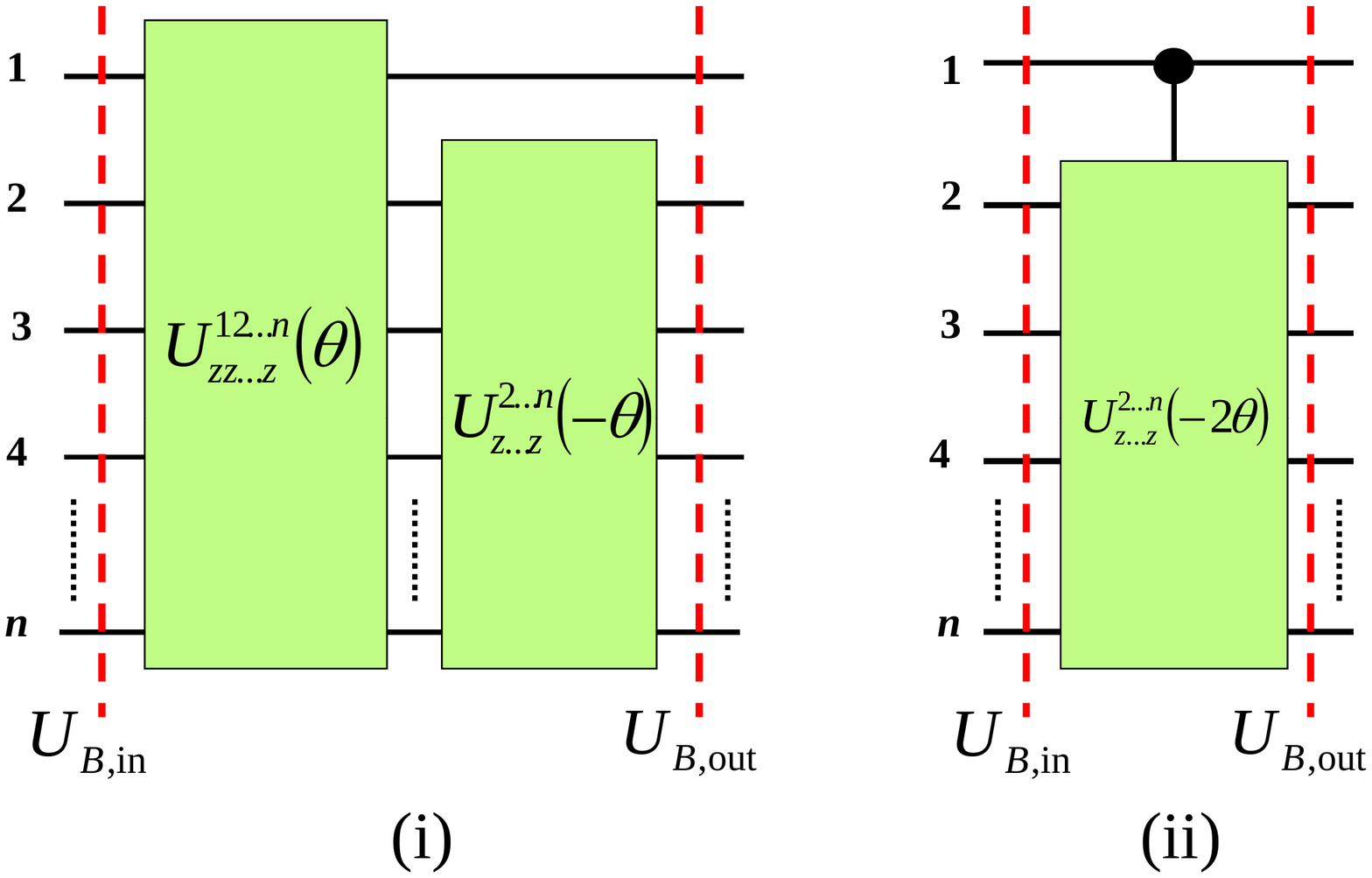}
 %l from the left, b from the bottom, r from the right, and t from the top. Where l, b, r and t are lengths.
\caption{(Color online) (i) The quantum circuit merely represents the temporal order of rotations for $\Lambda^{1}U^{2...n}_{z...z}(-2\theta)$. Horizontal lines stand for $n$ logical qubits. The first (green) rectangular box (from the left) symbolizes the $n$-qubit rotation $U^{12...n}_{zz...z}(\theta)$ and the second box symbolizes the $(n-1)$-qubit rotation $U^{2...n}_{z...z}(-\theta)$. Both of them are executed under the scheme described in Sec.~\ref{sec:Methodology1QC}. (ii) This circuit represents $\Lambda^{1}U^{2...n}_{z...z}(-2\theta)$, where qubit 1 is the control qubit and the other qubits are targets. The (red) dashed vertical lines at the input and the output section stand for the byproduct operators $\textit{\textbf{U}}_{B,\mathrm{in}}$ and $\textit{\textbf{U}}_{B,\mathrm{out}}$, respectively. Circuits~(i) and (ii) are equivalent.}\label{fig:2}
\end{figure}

The circuit representation of Eq.~(\ref{C1U}) is given in Fig.~\ref{fig:2}. In Fig.~\ref{fig:2}(i), the first and the second rectangular boxes depict $U^{12...n}_{zz...z}(\theta)$ (the first rotation) and $U^{2...n}_{z...z}(-\theta)$ (the second rotation), respectively. In practice, both of them are realized---by using a single ancilla qubit and a single measurement---with the methodology given in Sec.~\ref{sec:Methodology1QC} (see the example). Moreover, after executing the first rotation we bring back the ancilla qubit into an eigenstate of $X$ and use it for the second rotation. The implementation of $U^{12...n}_{zz...z}(\theta)$ and $U^{2...n}_{z...z}(-\theta)$ require $(1+n)$-qubit and $n$-qubit star graph states, respectively (see Fig.~\ref{fig:1}(i)), where the ancilla qubit is connected to the relevant logical qubits. The eigenvalues of the ancilla qubit corresponding to the first and the second rotation are $\kappa_{1}$ and $\kappa_{2}$, and the measurement outcomes are $m_{1}$ and $m_{2}$, respectively.

The classical information processing for this case has three main parts. The first part deals with the change in the measurement angles due to the byproduct operator $\textit{\textbf{U}}_{B,\mathrm{in}}$. $\textit{\textbf{U}}_{B,\mathrm{in}}$ appears just before implementing the first rotation and is denoted by the dashed vertical line at the input section in Fig.~\ref{fig:2}. When the gate $\Lambda^{1}U^{2...n}_{z...z}(-2\theta)$ in itself is a part of a circuit under simulation, then due to the implementation of previous gates the byproduct $\textit{\textbf{U}}_{B,\mathrm{in}}$ has emerged prior to the execution of $\Lambda^{1}U^{2...n}_{z...z}(-2\theta)$. Without loss of generality, we take $\textit{\textbf{U}}_{B,\mathrm{in}}=\prod^{n}_{j=1}(X^{(j)})^{x_{j}}(Z^{(j)})^{z_{j}}$ the same as given in Eq.~(\ref{I}). Only the $x$-part of the corresponding information flow vector $\mathcal{I}_{x,\mathrm{in}}$ influences the measurement bases $\left\{\arrowvert \uparrow, \downarrow (\pm\theta, \pi/2)\rangle_{a}\right\}$ for both rotations. According to Eq.~(\ref{HQCproU}), the angles $\theta$ for the first and $-\theta$ for the second rotation get altered. 
%%%%%
\begin{equation}
\begin{array}{lcl} %l=left, c=center, r=right.
\hphantom{-}\theta\rightarrow(-1)^{x}\theta & \mbox{for} & U^{12...n}_{zz...z}(\theta), \\
-\theta\rightarrow-(-1)^{x-x_{1}}\theta & \mbox{for} & U^{2...n}_{z...z}(-\theta), 
\end{array}
\label{thetax}
\end{equation} where $x$ is given by Eq.~(\ref{x}).
%%%%%

The second part deals with the eigenvalues $\kappa_{1}$ and $\kappa_{2}$, which influence the azimuthal angle $\pi/2$ of the measurement bases in the following way:
\begin{equation}
\begin{array}{lcl} %l=left, c=center, r=right.
\pi/2\rightarrow(-1)^{\kappa_{1}}\pi/2 & \mbox{for} & U^{12...n}_{zz...z}(\theta), \\
\pi/2\rightarrow(-1)^{\kappa_{2}}\pi/2 & \mbox{for} & U^{2...n}_{z...z}(-\theta). 
\end{array}
\label{phiK}
\end{equation}

The third part manages the contribution of measurement outcomes $m_{1}$ and $m_{2}$ to the byproduct operator $\textit{\textbf{U}}_{B,\textrm{in}}$. The implementation of both the first and the second rotation cause the byproduct operators $(Z^{\otimes n})^{m_{1}}$ and $(Z^{\otimes (n-1)})^{m_{2}}$ on the relevant logical qubits. Furthermore, these byproduct operators update $\textit{\textbf{U}}_{B,\textrm{in}}$ to $\textit{\textbf{U}}_{B,\textrm{out}}$. $\textit{\textbf{U}}_{B,\textrm{out}}$ is denoted by the dashed vertical line at the output section in Fig.~\ref{fig:2}. Only the $z$-part of the information flow vector $\mathcal{I}_{z,\textrm{in}}$ gets changed, while the $x$-part remains as it is, i.e., $\mathcal{I}_{x,\textrm{out}}=\mathcal{I}_{x,\textrm{in}}$,
\begin{equation}
\textit{\textbf{U}}_{B,\textrm{out}}=(X^{(1)})^{x_{1}}(Z^{(1)})^{z_{1}+m_{1}}\prod^{n}_{j=2}(X^{(j)})^{x_{j}}({Z}^{(j)})^{z_{j}+m_{1}+m_{2}}.
\end{equation}

%%%%%%%%%%%%%%%%%%%%%%%%%%%%%%%%%%%%%%%%%%%         FIGURE  (single control)                %%%%%%%%%%%%%%%%%%%%%%%%%%%%%%%%%%

%%%%%%%%%%%%%%%%%%%%%%%%%%%%%%%%%%%%%%%%%%%%%%%%%%%%%%%%%%%%%%%%%%%%%%%%%%%%%%%%%%%%%%%%%%%%%%%%%%%%%%%%%%%%%%%%%%%%%%%%%%%

%%%%%%%%%%%%%%%%%%%%%%%%%%%%%%%%%%%%%%%%%%%%%%%%%%%%%%%%%%%%%%%%%%%%%%%%%%%%%%%%%%%%%%%%%%%%%%%%%%%%%%%%%%%%%%%%%%%%%%%%%%%

%%%%%%%%%%%%%%%%%%%%%%%%%%%%%%%%%%%%%%%%%%%%           CASE (2): two-control         %%%%%%%%%%%%%%%%%%%%%%%%%%%%%%%%%%

\subsubsection{\label{sec:C2} The double-control gate $\Lambda^{12}U^{3...n}_{z...z}(4\theta)$}

\begin{figure}[t]
\centering
\includegraphics[ trim = 20mm 40mm 20mm 40mm, clip=true , width=\linewidth]{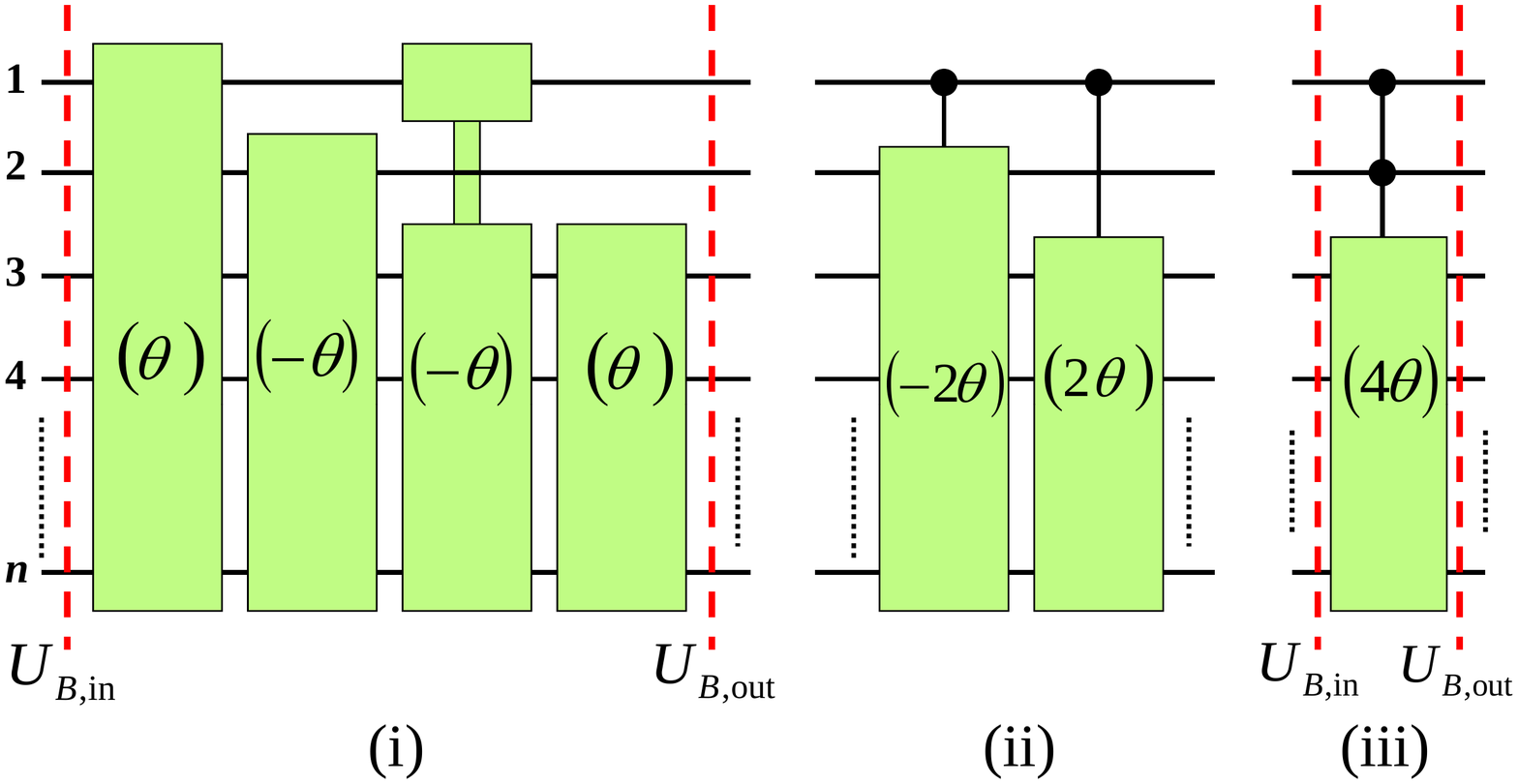}
 %l from the left, b from the bottom, r from the right, and t from the top. Where l, b, r and t are lengths.
\caption{(Color online) The horizontal lines stand for $n$ logical qubits. (i) Four (green) rectangular boxes (count from the left) represent $U^{12...n}_{zz...z}(\theta)$, $U^{2...n}_{z...z}(-\theta)$, $U^{13...n}_{z...z}(-\theta)$, and $U^{3...n}_{z...z}(\theta)$,  respectively. Each rotation is realized under the scheme described in Sec.~\ref{sec:Methodology1QC}. (ii) (from the left) The first (green) rectangular box symbolizes the $n$-qubit operation $\Lambda^{1}U^{2...n}_{z...z}(-2\theta)$, and the second one symbolizes the $(n-1)$-qubit operation $\Lambda^{1}U^{3...n}_{z...z}(2\theta)$. Both of them have the qubit 1 as control. (iii) The diagram represents $\Lambda^{12}U^{3...n}_{z...z}(4\theta)$, where the qubits 1 and 2 are the control qubits. In (i) and (iii), the (red) dashed vertical lines at the input and output section stand for the byproduct operators $\textbf{\textit{U}}_{B,\textrm{in}}$ and $\textbf{\textit{U}}_{B,\textrm{out}}$, respectively. Circuit~(ii) merely depicts the intermediate stage of circuits~(i) and (iii), and they all are mutually equivalent.}\label{fig:3}
\end{figure}

We have to combine two additional units $U^{13...n}_{z...z}(-\theta)$ (the third rotation) and $U^{3...n}_{z...z}(\theta)$ (the fourth rotation) with $\Lambda^{1}U^{2...n}_{z...z}(-2\theta)$ for the purpose of getting the gate $\Lambda^{12}U^{3...n}_{z...z}(4\theta)$.
In other words, $\Lambda^{12}U^{3...n}_{z...z}(4\theta)$ with two control qubits 1 and 2, is made up of four rotations, and its decomposition is given by
%%%%%%%%%%%%%%%%%%%%%%%%%%%%%%%%%%%%%%          C2U                 %%%%%%%%%%%%%%%%%%%%%%%%%%%%%%%%%%%%%%%%%%%%%%%%%%%%%%%%
\begin{equation}
\Lambda^{12}U^{3...n}_{z...z}(4\theta) =  U^{3...n}_{z...z}(\theta)U^{13...n}_{zz...z}(-\theta)U^{2...n}_{z...z}(-\theta) U^{12...n}_{zz...z}(\theta).\label{C2U}
\end{equation}
%\nonumber\\
%&=& \left(\arrowvert 00\rangle_{12}\langle 00\arrowvert + \arrowvert 01\rangle_{12}\langle 01\arrowvert+\arrowvert 10\rangle_{12}\langle 10\arrowvert \right)\otimes\textit{\textbf{I}}^{\otimes (n-2)}\nonumber \\
%& & +\arrowvert 11\rangle_{12}\langle 11\arrowvert \otimes U^{3...n}_{z...z}(4\theta)
%%%%%%%%%%%%%%%%%%%%%%%%%%%%%%%%%%%%%%%%%%%%%%%%%%%%%%%%%%%%%%%%%%%%%%%%%%%%%%%%%%%%%%%%%%%%%%%%%%%%%%%%%%%%%%%%%%%%%%%%%%%%% 
Figure~\ref{fig:3}(i) illustrates the temporal ordering of the multi-qubit rotations given in Eq.~(\ref{C2U}) by the rectangular boxes. 

The treatment for $\Lambda^{12}U^{3...n}_{z...z}(4\theta)$ is similar to that of Sec.~\ref{sec:C1}. All the rotations---$U^{12...n}_{zz...z}(\theta)$ (first), $U^{2...n}_{z...z}(-\theta)$ (second), $U^{13...n}_{z...z}(-\theta)$ (third), and $U^{3...n}_{z...z}(\theta)$ (fourth)---are performed one after another under the scheme given in Sec.~\ref{sec:Methodology1QC} (see the example). After initializing the ancilla qubit in the $X$ eigenbasis, we prepare a necessary star graph state for the first rotation, and then the ancilla qubit is measured in the appropriate basis. The measurement outcome is recorded, and the ancilla qubit is brought back again into an eigenstate of $X$ (recycled) for executing the next rotation. In this way, we can use the same ancilla qubit for all the rotations. $\kappa_{1}$, $\kappa_{2}$, $\kappa_{3}$, and $\kappa_{4}$ are the eigenvalues of the ancilla qubit, and $m_{1}$, $m_{2}$, $m_{3}$, and $m_{4}$ are the measurement outcomes corresponding to the first, second, third, and fourth rotation. As a side remark, one can also choose to perform these four rotations at the same time by using four different ancilla qubits, but this would require more hardware resources. We can take $\Lambda^{12}U^{3...n}_{z...z}(4\theta)$ as a single time step, because, in principle, all the four rotations can be executed at the same time.

%%%%%%%%%%%%%%%%%%%%%%%%%%%%%%%%%%%%%%%%%%%         FIGURE  (two-control)                %%%%%%%%%%%%%%%%%%%%%%%%%%%%%%%%%%

%%%%%%%%%%%%%%%%%%%%%%%%%%%%%%%%%%%%%%%%%%%%%%%%%%%%%%%%%%%%%%%%%%%%%%%%%%%%%%%%%%%%%%%%%%%%%%%%%%%%%%%%%%%%%%%%%%%%%%%%%%%%%%

The classical information processing for this case also has three main parts. The first part deals with the modification in the measurement angles because of the byproduct operator $\textit{\textbf{U}}_{B,\textrm{in}}=\prod^{n}_{i=1}(X ^{(j)})^{x_{j}}(Z^{(j)})^{z_{j}}$, which is represented by the dashed vertical line at the input section in Figs.~\ref{fig:3}(i) and \ref{fig:3}(iii). Here also, only $\mathcal{I}_{x,\textrm{in}}$ influences the measurement angle $\pm\theta$ for every rotation. 
%%%%%
\begin{equation}
\begin{array}{lcl} %l=left, c=center, r=right.
\hphantom{-}\theta\rightarrow(-1)^{x}\theta & \mbox{for} & U^{12...n}_{zz...z}(\theta), \\
-\theta\rightarrow-(-1)^{x-x_{1}}\theta & \mbox{for} & U^{2...n}_{z...z}(-\theta), \\
-\theta\rightarrow-(-1)^{x-x_{2}}\theta & \mbox{for} & U^{13...n}_{zz...z}(-\theta), \\
\hphantom{-}\theta\rightarrow(-1)^{x-x_{1}-x_{2}}\theta & \mbox{for} & U^{3...n}_{z...z}(\theta),
\end{array}
\label{Thetax}
\end{equation} 
where $x$ is the same given by Eq.~(\ref{x}).

%%%%%

The second part manages the influence of the eigenvalues $\kappa_{1}, \: \kappa_{2}, \: \kappa_{3},$ and $\kappa_{4}$ on the azimuthal angle $\pi/2$ of the measurement bases $\left\{\arrowvert \uparrow, \downarrow (\pm\theta, \pi/2)\rangle_{a}\right\}$ in the following way: 
%%%%%
\begin{equation}
\begin{array}{lcl} %l=left, c=center, r=right.
\pi/2\rightarrow(-1)^{\kappa_{1}}\pi/2 & \mbox{for} & U^{12...n}_{zz...z}(\theta), \\
\pi/2\rightarrow(-1)^{\kappa_{2}}\pi/2 & \mbox{for} & U^{2...n}_{z...z}(-\theta), \\
\pi/2\rightarrow(-1)^{\kappa_{3}}\pi/2 & \mbox{for} & U^{13...n}_{zz...z}(-\theta), \\
\pi/2\rightarrow(-1)^{\kappa_{4}}\pi/2 & \mbox{for} & U^{3...n}_{z...z}(\theta).
\end{array}\label{PhiK}
\end{equation}
%%%%%%%%%%%%%
%%%%%%%%%%%%%%%%%%%%%%%%%%%%%%%%%%%%%%%%%%%%%%%%%%%%%%%%%%%%%%%%%%%%%%%%%%%%%%%%%%%%%%%%%%%%%%%%%%%%%%%%%%%%%%%%%%%%%%%%%%%%

%%%%%%%%%%%%%%%%%%%%%%%%%%%%%%%%%%%%%%%%%%%%%%%%%%%%% FIGURE-4 three control  %%%%%%%%%%%%%%%%%%%%%%%%%%%%%%%%%%%%%%%%%%%%%%%

%%%%%%%%%%%%%%%%%%%%%%%%%%%%%%%%%%%%%%%%%%%%%%%%%%%%%%%%%%%%%%%%%%%%%%%%%%%%%%%%%%%%%%%%%%%%%%%%%%%%%%%%%%%%%%%%%%%%%%%%%%%%

\begin{figure*}[t] %htbp! \here", \top", \bottom", \page", and \as soon as possible", respectively
\centering %A bad point of figure* environment is that they can be placed only at the top of the page or on their own page.
\includegraphics[ trim = 25mm 50mm 25mm 50mm, clip=true , width=0.7\linewidth]{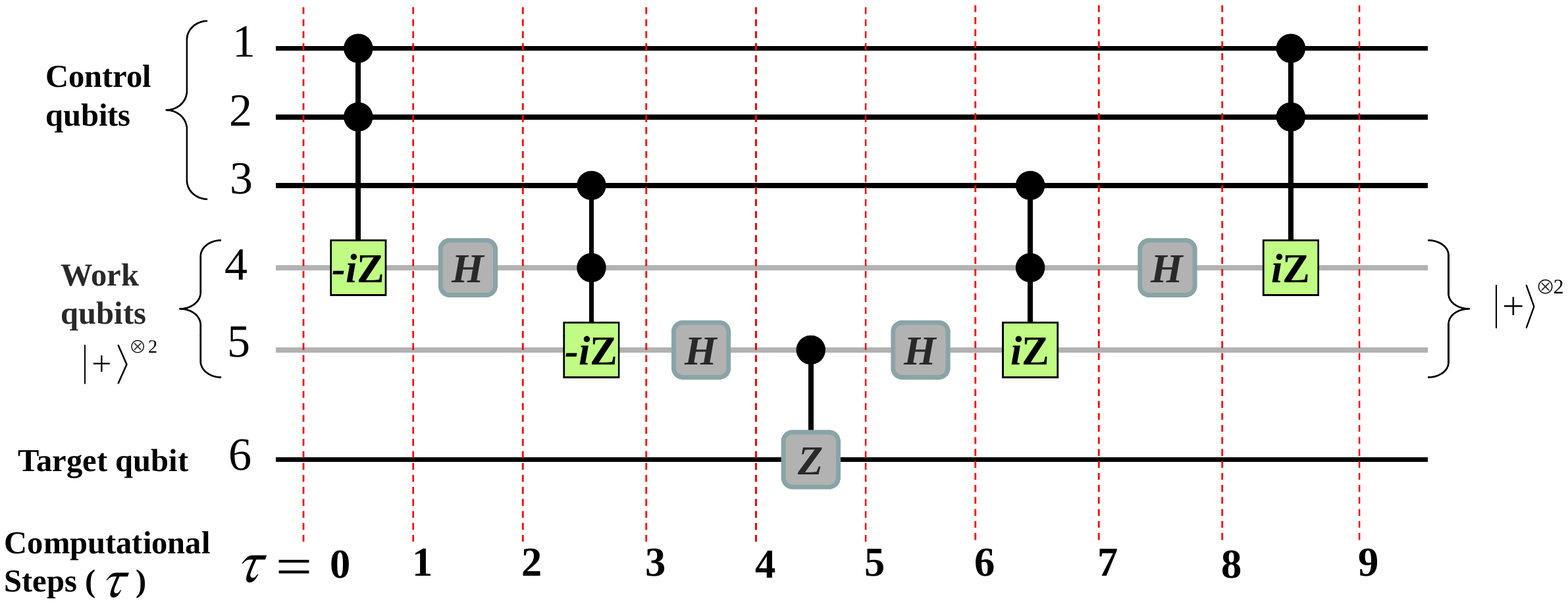}
 %l from the left, b from the bottom, r from the right, and t from the top. Where l, b, r and t are lengths.
\caption{(Color online) The quantum circuit for implementing the 4-qubit gate $\Lambda^{123}_{(3)}Z^{6}$. From the top, three black horizontal lines stand for the control qubits 1, 2, and 3, and the next two gray horizontal lines stand for the work qubits 4 and 5, which are prepared in the state $\arrowvert +\rangle^{\otimes2}$. The black horizontal line at the bottom stands for the target qubit 6. Every 3-qubit gate is the special case of $\Lambda^{12}U^{3...n}_{z...z}(4\theta)$ (for $n =3$ and $\theta=\pm\pi/4$), and they are realized by the four rotations according to Fig.~\ref{fig:3}. The Hadamard gate $H$ and $\textsc{cz}(5,6)$ are executed by the unitary evolution. The computation steps ($\tau$) are expressed by (red) dashed vertical lines for the classical information processing according to Table~\ref{table:CIP}.}\label{fig:4}
\end{figure*}

%%%%%%%%%%%%%%%%%%%%%%%%%%%%%%%%%%%%%%%%%%%%%%%%%%%%%%%%%%%%%%%%%%%%%%%%%%%%%%%%%%%%%%%%%%%%%%%%%%%%%%%%%%%%%%%%%%%%%%%%%%

%%%%%%%%%%%%%%%%%%%%%%%%%%%%%%%%%%%%%%%%%%%%%%%%%%%%%%%%%%%%%%%%%%%%%%%%%%%%%%%%%%%%%%%%%%%%%%%%%%%%%%%%%%%%%%%%%%%%%%%%%%%%

The third part handles the random measurement outcomes $m_{1}$, $m_{2}$, $m_{3}$, and $m_{4}$; which cause the byproduct operators $\left(Z^{\otimes n}\right)^{m_{1}}$, $\left(Z^{\otimes (n-1)}\right)^{m_{2}}$, $\left(Z^{\otimes (n-1)}\right)^{m_{3}}$, and $\left(Z^{\otimes (n-2)}\right)^{m_{4}}$, respectively, on the relevant logical qubits. Furthermore, they change $\textit{\textbf{U}}_{B,\textrm{in}}$ into $\textit{\textbf{U}}_{B,\textrm{out}}$ by their contribution. $\textit{\textbf{U}}_{B,\textrm{out}}$ is denoted by the dashed vertical line at the output section in Figs.~\ref{fig:3}(i) and \ref{fig:3}(iii). So, only the $z$-part of the corresponding information flow vector $\mathcal{I}_{z,\textrm{in}}$ gets changed to $\mathcal{I}_{z,\textrm{out}}$, while $\mathcal{I}_{x,\textrm{out}}=\mathcal{I}_{x,\textrm{in}}$.
\begin{equation}
\mathcal{I}_{z,\textrm{out}}=\left(
\begin{array}{cc}
z_{1}+m_{1}+m_{3}	\\
z_{2}+m_{1}+m_{2} \\
z_{3}+m	\\
\vdots\\
z_{n}+m
\end{array}
\right), \label{Ivector}	
\end{equation}
where $$m = m_{1}+m_{2}+m_{3}+m_{4}.$$

We emphasize two points. First, $\Lambda^{12}U^{3...n}_{z...z}(4\theta)$ is symmetric under permutation of the qubits. So, we can take any two qubits as controls and the rest of the qubits as targets by using only four multi-qubit rotations. If we continue along the same direction, then the controlled rotation $\Lambda^{1...c}U^{(c+1)...n}_{z...z}(\theta)$ with $c$  control qubits requires $2^{c}$ units of rotation. The number $2^{c}$ is independent of the number of target qubits $(n-c)$, but when $c$ becomes order of $n$, then $2^{c}$ becomes exponential in $n$. In order to fix this exponential growth problem, we need some extra work qubits \cite{Barenco95}. Our next example will justify this remark. 

Second, $\Lambda^{12}U^{3...n}_{z...z}(4\theta)$ becomes $\Lambda^{12}U^{3}_{z}(4\theta)$ for $n=3$. The gate $\Lambda^{12}U^{3}_{z}(4\theta)$ is equivalent to the Deutsch's universal gate $\Lambda^{12}\left[iR^{3}_{x}(4\theta)\right]$ \cite{Deutsch89} up to single-qubit gates, provided the angle `$4\theta$' is incommensurate with $\pi$. We are going to use $\Lambda^{12}U^{3}_{z}(\pm\pi)$ with two work qubits for implementing our next gate $\Lambda^{123}Z^{6}$.  
%%%%%%%%%%%%%%%%%%%%%%%%%%%%%%%%%%%%%%%%%%%%%%%%%%%%%%%%%%%%%%%%%%%%%%%%%%%%%%%%%%%%%%%%%%%%%%%%%%%%%%%%%%%%%%%%%%%%%%%%%%%%

%%%%%%%%%%%%%%%%%%%%%%%%%%%%%%%%%%%%%%%%%%%%%%%            CASE (3): three-control         %%%%%%%%%%%%%%%%%%%%%%%%%%%%%%%%%%

%%%%%%%%%%%%%%%%%%%%%%%%%%%%%%%%%%%%%%%%%%%%%%%%%%%%%%%%%%%%%%%%%%%%%%%%%%%%%%%%%%%%%%%%%%%%%%%%%%%%%%%%%%%%%%%%%%%%%%%%%%%%
\subsubsection{\label{sec:C3} The triple-control gate $\Lambda^{123}Z^{6}$}

We put the bits and pieces of the HQCM together as a summary by taking the gate $\Lambda^{123}Z^{6}$ as an example. The complete scheme about its implementation in terms of its circuit diagram is shown in Fig.~\ref{fig:4}, and the associated classical information-processing parts are given in Table~\ref{table:CIP}.

First, we efficiently decompose the gate $\Lambda^{123}Z^{6}$ into a sequence of elementary gates (single-qubit gates, the $\textsc{cz}$ gate, and $U^{12...n}_{zz...z}(\theta)$). The temporal order of the elementary gates for $\Lambda^{123}Z^{6}$ is depicted by the circuit diagram in Fig.~\ref{fig:4}, where the qubits 1, 2, and 3 act as the control qubits and the qubit 6 as the target qubit, and they all are represented by the black horizontal lines. The work qubits 4 and 5 are initialized in the state $\arrowvert +\rangle^{\otimes 2}$, and they are represented by the gray horizontal lines in the figure. The work qubits are used to make the decomposition of $\Lambda^{123}Z^{6}$ economical.

The 3-qubit gates $\Lambda^{12}U^{4}_{z}(\pi)$ (1st gate), $\Lambda^{34}U^{5}_{z}(\pi)$ (3rd gate), $\Lambda^{34}U^{5}_{z}(-\pi)$ (7th gate), and $\Lambda^{12}U^{4}_{z}(-\pi)$ (9th gate) are represented by rectangular boxes with double-control. Every 3-qubit gate is further decomposed into four rotations around the $z$-axis according to Eq.~(\ref{C2U}), here $n =3$ and $\theta=\pm\pi/4$. Furthermore, each rotation is executed by preparing a required star graph state, followed by the measurement in the appropriate basis. The detailed methodology is already mentioned in Sec.~\ref{sec:C2}. The Hadamard gates $H$ are displayed by rounded rectangles, and the two-qubit gate $\textsc{cz}(5,6)$ is shown by the rounded rectangle on the qubit 6 with the qubit 5 as control in Fig.~\ref{fig:4}. The Hadamard and the $\textsc{cz}(5,6)$ gates are executed by the unitary evolution.

The classical information-processing parts for $\Lambda^{123}Z^{6}$ are handled by a classical computer according to Table~\ref{table:CIP}. In this table, the first column is for the computational steps $\tau$, which are represented by the dashed vertical lines in Fig.~\ref{fig:4}. There are ten vertical lines in the figure and ten rows in the table for the ten computational steps from $\textbf{0}$ to $\textbf{9}$. At each vertical line the information flow vector $\mathcal{I}(\tau)$ gets updated. The second and the third columns are reserved for $\mathcal{I}_{x}(\tau)$ and $\mathcal{I}_{z}(\tau)$ respectively. If required, the change in the measurement angles for the next gate based on the updated value of $\mathcal{I}(\tau)$ is calculated; they are given in the fourth column. After performing the measurements in the appropriate bases, the measurement outcomes are recorded in the fifth column. 
%%%%%%%%%%%%%%%%%%%%%%%%%%%%%%%%%%%%%%%%%%%%%%%%%%%%%%%%%%%%%%%%%%%%%%%%%%%%%%%%%%%%%%%%%%%%%%%%%%%%%%%%%%%%%%%%%%%%%%%%%%%%%%

%%%%%%%%%%%%%%%%%%%%%%%%%%%%%%%%%%%%%%%%%%%%%%%%%%%%%%%%%%%%%%%%%%%%%%%%%%%%%%%%%%%%%%%%%%%%%%%%%%%%%%%%%%%%%%%%%%%%%%%%%%%%%%

%                                                 TABLE

%%%%%%%%%%%%%%%%%%%%%%%%%%%%%%%%%%%%%%%%%%%%%%%%%%%%%%%%%%%%%%%%%%%%%%%%%%%%%%%%%%%%%%%%%%%%%%%%%%%%%%%%%%%%%%%%%%%%%%%%%%%%%

%%%%%%%%%%%%%%%%%%%%%%%%%%%%%%%%%%%%%%%%%%%%%%%%%%%%%%%%%%%%%%%%%%%%%%%%%%%%%%%%%%%%%%%%%%%%%%%%%%%%%%%%%%%%%%%%%%%%%%%%%%%%%%

\begingroup

\squeezetable % squeeze the table
\begin{table*}[t]% \table* environment will place the table across both columns (the table usually will appear either at the top or the bottom of the following page).
\begin{ruledtabular}
\caption{The classical information-processing parts for $\Lambda^{123}Z^{6}$} % title of Table
\centering % used for centering table
    \begin{tabular*}{\textwidth}{@{\extracolsep{\fill}}  c  c  c  p{5cm} p{5cm} }

    $\tau$ & $\mathcal{I}_{x}(\tau)$ & $\mathcal{I}_{z}(\tau)$ & Angle $\pm\theta$ (here $\theta=\pi/4$) & Measurement outcomes\\ \hline
    
    \textbf{0} & $\left(\begin{matrix}0\\0\\0\\0\\0\\0\end{matrix}\right)$ & 
$\left(\begin{matrix}0\\0\\0\\0\\0\\0\end{matrix}\right)$ & 
$\begin{array}{lcl} %l=left, c=center, r=right.
\mbox{No change in angle} & \mbox{for} & U^{124}_{zzz}(\theta) \\
\mbox{No change in angle} & \mbox{for} & U^{24}_{zz}(-\theta) \\
\mbox{No change in angle} & \mbox{for} & U^{14}_{zz}(-\theta) \\
\mbox{No change in angle} & \mbox{for} & U^{4}_{z}(\theta)
\end{array}$ & 
$\begin{array}{lcl} %l=left, c=center, r=right.
m_{11} & \mbox{for} & U^{124}_{zzz}(\theta) \\
m_{12} & \mbox{for} & U^{24}_{zz}(-\theta) \\
m_{13} & \mbox{for} & U^{14}_{zz}(-\theta) \\
m_{14} & \mbox{for} & U^{4}_{z}(\theta)
\end{array}$ 

$ m_{1} = m_{11}+m_{12}+m_{13}+m_{14}$ \\ 

    \textbf{1} & $\left(\begin{matrix}0\\0\\0\\0\\0\\0\end{matrix}\right)$ &     
$\left(\begin{matrix}m_{11}+m_{13}\\m_{11}+m_{12}\\0\\m_{1}\\0\\0\end{matrix}\right)$ &  &  \\ 

    \textbf{2} & $\left(\begin{matrix}0\\0\\0\\m_{1}\\0\\0\end{matrix}\right)$ &     
$\left(\begin{matrix}m_{11}+m_{13}\\m_{11}+m_{12}\\0\\0\\0\\0\end{matrix}\right)$ & 
$\begin{array}{lcl} %l=left, c=center, r=right.
\hphantom{-}\theta\rightarrow(-1)^{m_{1}}\theta & \mbox{for} & U^{345}_{zzz}(\theta) \\
-\theta\rightarrow-(-1)^{m_{1}}\theta & \mbox{for} & U^{45}_{zz}(-\theta) \\
\mbox{No change in angle} & \mbox{for} & U^{35}_{zz}(-\theta) \\
\mbox{No change in angle} & \mbox{for} & U^{5}_{z}(\theta)
\end{array}$ & 
$\begin{array}{lcl} %l=left, c=center, r=right.
m_{31} & \mbox{for} & U^{345}_{zzz}(\theta) \\
m_{32} & \mbox{for} & U^{45}_{zz}(-\theta) \\
m_{33} & \mbox{for} & U^{35}_{zz}(-\theta) \\
m_{34} & \mbox{for} & U^{5}_{z}(\theta)
\end{array}$ 

$ m_{3} = m_{31}+m_{32}+m_{33}+m_{34}$ \\ 

    \textbf{3} & $\left(\begin{matrix}0\\0\\0\\m_{1}\\0\\0\end{matrix}\right)$ &     
$\left(\begin{matrix}m_{11}+m_{13}\\m_{11}+m_{12}\\m_{31}+m_{33}\\m_{31}+m_{32}\\m_{3}\\0\end{matrix}\right)$ &  &  \\ 

    \textbf{4} & $\left(\begin{matrix}0\\0\\0\\m_{1}\\m_{3}\\0\end{matrix}\right)$ &     
$\left(\begin{matrix}m_{11}+m_{13}\\m_{11}+m_{12}\\m_{31}+m_{33}\\m_{31}+m_{32}\\0\\0\end{matrix}\right)$ &  &  \\ 

    \textbf{5} & $\left(\begin{matrix}0\\0\\0\\m_{1}\\m_{3}\\0\end{matrix}\right)$ &     
$\left(\begin{matrix}m_{11}+m_{13}\\m_{11}+m_{12}\\m_{31}+m_{33}\\m_{31}+m_{32}\\0\\m_{3}\end{matrix}\right)$ &  &  \\ 

    \textbf{6} & $\left(\begin{matrix}0\\0\\0\\m_{1}\\0\\0\end{matrix}\right)$ &     
$\left(\begin{matrix}m_{11}+m_{13}\\m_{11}+m_{12}\\m_{31}+m_{33}\\m_{31}+m_{32}\\m_{3}\\m_{3}\end{matrix}\right)$ & $\begin{array}{lcl} %l=left, c=center, r=right.
-\theta\rightarrow-(-1)^{m_{1}}\theta & \mbox{for} & U^{345}_{zzz}(-\theta) \\
\hphantom{-}\theta\rightarrow(-1)^{m_{1}}\theta & \mbox{for} & U^{45}_{zz}(\theta) \\
\mbox{No change in angle} & \mbox{for} & U^{35}_{zz}(\theta) \\
\mbox{No change in angle} & \mbox{for} & U^{5}_{z}(-\theta)
\end{array}$ & 
$\begin{array}{lcl} %l=left, c=center, r=right.
m_{71} & \mbox{for} & U^{345}_{zzz}(-\theta) \\
m_{72} & \mbox{for} & U^{45}_{zz}(\theta) \\
m_{73} & \mbox{for} & U^{35}_{zz}(\theta) \\
m_{74} & \mbox{for} & U^{5}_{z}(-\theta)
\end{array}$ 

$ m_{7} = m_{71}+m_{72}+m_{73}+m_{74}$ \\ 

    \textbf{7} & $\left(\begin{matrix}0\\0\\0\\m_{1}\\0\\0\end{matrix}\right)$ &     
$\left(\begin{matrix}m_{11}+m_{13}\\m_{11}+m_{12}\\m_{31}+m_{33}+m_{71}+m_{73}\\m_{31}+m_{32}+m_{71}+m_{72}\\m_{3}+m_{7}\\m_{3}\end{matrix}\right)$ &  & \\ 

    \textbf{8} & $\left(\begin{matrix}0\\0\\0\\m_{31}+m_{32}+m_{71}+m_{72}\\0\\0\end{matrix}\right)$ &     
$\left(\begin{matrix}m_{11}+m_{13}\\m_{11}+m_{12}\\m_{31}+m_{33}+m_{71}+m_{73}\\m_{1}\\m_{3}+m_{7}\\m_{3}\end{matrix}\right)$ & $\begin{array}{lcl} %l=left, c=center, r=right.
-\theta\rightarrow-(-1)^{\tilde{m}}\theta & \mbox{for} & U^{124}_{zzz}(-\theta) \\
\hphantom{-}\theta\rightarrow(-1)^{\tilde{m}}\theta & \mbox{for} & U^{24}_{zz}(\theta) \\
\hphantom{-}\theta\rightarrow(-1)^{\tilde{m}}\theta & \mbox{for} & U^{14}_{zz}(\theta) \\
-\theta\rightarrow-(-1)^{\tilde{m}}\theta & \mbox{for} & U^{4}_{z}(-\theta)
\end{array}$ 

$ \tilde{m} = m_{31}+m_{32}+m_{71}+m_{72}$ & 
$\begin{array}{lcl} %l=left, c=center, r=right.
m_{91} & \mbox{for} & U^{124}_{zzz}(-\theta) \\
m_{92} & \mbox{for} & U^{24}_{zz}(\theta) \\
m_{93} & \mbox{for} & U^{14}_{zz}(\theta) \\
m_{94} & \mbox{for} & U^{4}_{z}(-\theta)
\end{array}$ 

$ m_{9} = m_{91}+m_{92}+m_{93}+m_{94}$ \\ 

    \textbf{9} & $\left(\begin{matrix}0\\0\\0\\m_{31}+m_{32}+m_{71}+m_{72}\\0\\0\end{matrix}\right)$ &     
$\left(\begin{matrix}m_{11}+m_{13}+m_{91}+m_{93}\\m_{11}+m_{12}+m_{91}+m_{92}\\m_{31}+m_{33}+m_{71}+m_{73}\\m_{1}+m_{9}\\m_{3}+m_{7}\\m_{3}\end{matrix}\right)$  \\ 

\end{tabular*}\label{table:CIP}
\end{ruledtabular}
\end{table*}
\endgroup

%%%%%%%%%%%%%%%%%%%%%%%%%%%%%%%%%%%%%%%%%%%%%%%%%%%%%%%%%%%%%%%%%%%%%%%%%%%%%%%%%%%%%%%%%%%%%%%%%%%%%%%%%%%%%%%%%%%%%%%%%%%%%

%%%%%%%%%%%%%%%%%%%%%%%%%%%%%%%%%%%%%%%%%%%%%%%%%%%%%%%%%%%%%%%%%%%%%%%%%%%%%%%%%%%%%%%%%%%%%%%%%%%%%%%%%%%%%%%%%%%%%%%%%%%%%%

%%%%%%%%%%%%%%%%%%%%%%%%%%%%%%%%%%%%%%%%%%%%%%%%%%%%%%%%%%%%%%%%%%%%%%%%%%%%%%%%%%%%%%%%%%%%%%%%%%%%%%%%%%%%%%%%%%%%%%%%%%%%%%

Let us go through the table row by row. Before starting the computation (in the first row $\tau=\textbf{0}$), all the entries of both $\mathcal{I}_{x}(\textbf{0})$ and $\mathcal{I}_{z}(\textbf{0})$ are zeros (initialization). So, there is no change in the measurement angle for each of the four rotations associated with the gate $\Lambda^{12}U^{4}_{z}(\pi)$ (1st gate). The measurement outcomes $m_{11}$, $m_{12}$, $m_{13}$, and $m_{14}$ corresponding to the four rotations $U^{124}_{zzz}(\theta)$, $U^{24}_{zz}(-\theta)$, $U^{14}_{zz}(-\theta)$, and $U^{4}_{z}(\theta)$ are recorded. These outcomes give some nonzero entries to $\mathcal{I}_{z}(\textbf{1})$ according to Eq.~(\ref{Ivector}). The measurement outcome $m_{jk}$ corresponds to the $k$th rotation of the $j$th 3-qubit gate. The next gate in the circuit is the Hadamard gate $H^{(4)}$, which does not change under the propagation of the byproduct operator; that is why the forth column of the second row $\tau=\textbf{1}$ is empty. The $H$-gate is realized by the unitary evolution, that is why the fifth column of the second row is also empty. But the $H$-gate changes the information flow vector $\mathcal{I}(\textbf{1})$ into $\mathcal{I}(\textbf{2})$ under the propagation relation given by Eq.~(\ref{proH}), and the propagation matrix for the $H$-gate is given by Eqs.~(\ref{CH}) and (\ref{CCH}). The $3$rd gate is $\Lambda^{34}U^{5}_{z}(\pi)$. The measurement angles $\pm\theta$ only for the rotation $U^{345}_{zzz}(\theta)$ and $U^{45}_{zz}(-\theta)$ get influenced by $\mathcal{I}_{x}(\textbf{2})$ according to Eq.~(\ref{Thetax}). The measurement outcomes $m_{31}$, $m_{32}$, $m_{33}$, and $m_{34}$ only transform $\mathcal{I}_{z}(\textbf{2})$ into $\mathcal{I}_{z}(\textbf{3})$. In this way going through Table~\ref{table:CIP} along with Fig.~\ref{fig:4} explains the whole scheme, and the final output result is interpreted according to Eq.~(\ref{sx}) with the help of $\mathcal{I}_{x}(\textbf{9})$. 

Here, the $z$-part of the information flow vector $\mathcal{I}_{z}(\tau)$ gets the new entries from the implementation of 3-qubit gates only according to Eq.~(\ref{Ivector}). The entries of $\mathcal{I}(\tau)$ get manipulated under the propagation of the byproduct operator through the Hadamard gates and the $\textsc{cz}(5,6)$ gate according to the propagation relations~(\ref{proH}) and (\ref{HQCproCZ}), respectively. But the propagation of the byproduct operator does not change the $H$- and $\textsc{cz}$-gate. The $x$-part of the information flow vector $\mathcal{I}_{x}(\tau)$ influences the measurement angles $\pm\theta$ of the rotations for every 3-qubit gate according to Eq.~(\ref{Thetax}). As a side remark, the sign of the azimuthal angle $\pi/2$ of the measurement bases for the rotations also depends on the eigenvalues of the ancilla qubit according to Eq.~(\ref{PhiK}), which is not mentioned in the table.
 
We can easily generalize this example up the $n$-qubit gate $\Lambda^{12...(n-1)}Z^{n}$. Where the $n-1$ logical qubits $1,2..., n-1$ are the control qubits and the last one is the target qubit. For implementing this gate we need $n-2$ work qubits, which are initialized in the state $\arrowvert +\rangle^{\otimes(n-2)}$. In Grover's search algorithm this gate plays a very important role, which we are discussing in the next section.

%%%%%%%%%%%%%%%%%%%%%%%%%%%%%%%%%%%%%%%%%%%%%%%%%%%%%%%%%%%%%%%%%%%%%%%%%%%%%%%%%%%%%%%%%%%%%%%%%%%%%%%%%%%%%%%%%%%%%%%%%%%%%%%%%%%%

%%%%%%%%%%%%%%%%%%%%%%%%%%%%%%%%%%%%%%%%%%%                                                %%%%%%%%%%%%%%%%%%%%%%%%%%%%%%%%%%%%%%%%%
%%%%%%%%%%%%%%%%%%%%%%%%%%%%%%%%%%%%%%%%%%%                 GA                            %%%%%%%%%%%%%%%%%%%%%%%%%%%%%%%%%%%%%%%%%%
%%%%%%%%%%%%%%%%%%%%%%%%%%%%%%%%%%%%%%%%%%%                                                %%%%%%%%%%%%%%%%%%%%%%%%%%%%%%%%%%%%%%%%%

%%%%%%%%%%%%%%%%%%%%%%%%%%%%%%%%%%%%%%%%%%%%%%%%%%%%%%%%%%%%%%%%%%%%%%%%%%%%%%%%%%%%%%%%%%%%%%%%%%%%%%%%%%%%%%%%%%%%%%%%%%%%%%%%%%%%

\section{\label{sec:GA}Grover's Algorithm within the HQCM}
In this section, we illustrate the use of the HQCM in the context of a practical example: Grover's search algorithm (GA) \cite{Grover97}. Here, we shall see that the $n$-qubit gate $\Lambda^{12...(n-1)}Z^{n}$ along with the single-qubit operations is sufficient to perform GA.

The best known quantum algorithm, for an unstructured database search, is GA. We have an unstructured database of total $\textbf{N}$ items, out of which only one item matches with our query. Where a classical computer takes an average $\Theta(\textbf{N}/2)$ steps, a QC with GA takes only $\Theta(\sqrt{\textbf{N}})$ steps (Grover's iterations) to find out the marked item. A ``step'' is a query of the oracle in the current context.

A brief description of GA is given as follows: We can recognize each item of our database by a $n$-bit string. For simplicity, we choose $\textbf{N}=2^{n}$ = the total number of $n$-bit strings, and only one string out of $\textbf{N}$ is marked. The task of the search problem is to recover the marked $n$-bit string in the end of computation. GA begins with the Hadamard operation on each qubit, which were initially prepared in the state $\arrowvert 0\rangle^{\otimes n}$, and that creates the superposition of the kets for all possible $n$-bit strings with equal amplitude. 

The next step is the implementation of Grover's iteration, which is a rotation in effect. It can be decomposed into two reflection operations. The first reflection operator is the oracle $\mathcal{O}$, which has the ability to recognize the solution of the search problem. Mathematically, $\mathcal{O}$ gives a conditional phase shift of $\pi$ to the matching string only. The second reflection operator is the diffusion operator $\mathcal{D}$, which gives an inversion about the average. Like other quantum algorithms, GA is also probabilistic in nature. After $\Theta(\sqrt{\textbf{N}})$ steps (iterations) the amplitude of the marked string becomes significantly larger than those of the unmarked strings. Finally, we read the output by performing measurements on all the qubits. There have been many successful attempts at the implementation of GA, for $\textbf{N}=4$ or $\textbf{N}=8$, in different physical setups such as with NMR system \cite{nmrGA1, nmrGA2, nmrGA3}, with cavity quantum electrodynamics (QED) \cite{qedGA1, qedGA2, qedGA3}, with optics \cite{opticsGA1, opticsGA2}, and with the MQCM \cite{mbqcGA1, mbqcGA2}. 

Now, let us take a look at the structure of these two reflection operators. If only one out of $\textbf{N}$ item matches with our query; then there exist $\textbf{N}$ different $\mathcal{O}$ operations (one for each item). Mathematically, the oracle corresponds to the $j$th item can be written as 
\begin{equation}
\mathcal{O}^{j}=I^{\otimes n}-2\arrowvert j \rangle \langle j \arrowvert.
\label{Oj} 
\end{equation}
The case of $j=\textbf{N}$, $\mathcal{O}^{\textbf{N}}=I^{\otimes n}-2\arrowvert \textbf{N}\rangle \langle \textbf{N}\arrowvert$, corresponds to $\arrowvert \textbf{N}\rangle=\arrowvert 1\rangle^{\otimes n}$ and is nothing but the $n$-qubit gate $\Lambda^{12...(n-1)}Z^{n}$. We already discussed its implementation with the HQCM for $n=4$ in Sec.~\ref{sec:C3}, and its circuit diagram is shown in Fig.~\ref{fig:4}. Any other oracle can be derived by performing the gate(s) $\textbf{\textit{X}}$ on the relevant qubit(s) before and after performing the gate $\Lambda^{12...(n-1)}Z^{n}$. For example, the oracle associated with the item $\arrowvert 1\rangle=\arrowvert 0\rangle^{\otimes n}$ can be derived as 
\begin{equation}
\mathcal{O}^{1}=X^{\otimes n}\left[\Lambda^{12...(n-1)}Z^{n}\right]X^{\otimes n}. 
\label{O1} 
\end{equation}
So the gate $\Lambda^{12...(n-1)}Z^{n}$ is used for implementing every oracle. The information, ``which of the oracle is executed by the black box?", is hidden to us. In other words, we do not know on which qubit(s) the black box is implementing the $X$ gate(s) along with $\Lambda^{12...(n-1)}Z^{n}$.

The mathematical structure of the diffusion operator is 
\begin{equation}
\mathcal{D}=-H^{\otimes n}X^{\otimes n}\left[\Lambda^{12...(n-1)}_{(n-1)}Z^{n}\right]X^{\otimes n}H^{\otimes n}.
\label{D} 
\end{equation}
It can also be constructed by performing the Hadamard- and the $X$-gate on every logical qubit before and after performing the gate $\Lambda^{12...(n-1)}Z^{n}$. In summary, the gate $\Lambda^{12...(n-1)}Z^{n}$ together with the Hadamard and the $X$ gates is sufficient to realize GA in the framework of HQCM.

%%%%%%%%%%%%%%%%%%%%%%%%%%%%%%%%%%%%%%%%%%%%%%%%%%%%%%%%%%%%%%%%%%%%%%%%%%%%%%%%%%%%%%%%%%%%%%%%%%%%%%%%%%%%%%%%%%%%%%%%%%%%%%%%%%%%

%%%%%%%%%%%%%%%%%%%%%%%%%%%%%%%%%%%%%%%%%%%                                                %%%%%%%%%%%%%%%%%%%%%%%%%%%%%%%%%%%%%%%%%
%%%%%%%%%%%%%%%%%%%%%%%%%%%%%%%%%%%%%%%%%%%                 Conclusion                            %%%%%%%%%%%%%%%%%%%%%%%%%%%%%%%%%%%%%%%%%%
%%%%%%%%%%%%%%%%%%%%%%%%%%%%%%%%%%%%%%%%%%%                                                %%%%%%%%%%%%%%%%%%%%%%%%%%%%%%%%%%%%%%%%%

%%%%%%%%%%%%%%%%%%%%%%%%%%%%%%%%%%%%%%%%%%%%%%%%%%%%%%%%%%%%%%%%%%%%%%%%%%%%%%%%%%%%%%%%%%%%%%%%%%%%%%%%%%%%%%%%%%%%%%%%%%%%%%%%%%%%

\section{\label{sec:Conc}Summary and outlook}

We have established the HQCM, which is a hybrid model of the MQCM and the UQCM, where at first, a big unitary gate under simulation is decomposed into a sequence of elementary gates. The elementary gates in the HQCM are an arbitrary single-qubit gate, the $\textsc{cz}$ gate and the multi-qubit rotation around the $z$-axis. Every single-qubit gate and the $\textsc{cz}$-gate are realized by a unitary evolution. Every multi-qubit rotation is executed by preparing a respective star graph state followed by a single measurement. The HQCM is a model where a portion of the quantum circuit is simulated by the unitary evolution, and the rest is by the measurements. 

The choice of elementary gates is governed by the experimental easiness in terms of resources and computational steps. The implementation of an arbitrary single-qubit gate with the unitary evolution is straightforward in comparison with its implementation with the MQCM. The $\textsc{cz}$-gate in itself is the part of experimental setup for creating the graph states. The star graph states for the multi-qubit rotations can be realized in one shot, and a single measurement is enough for executing these rotations. 

The classical information processing in the HQCM is very simple in comparison with the MQCM. In the HQCM, only the $2n$-component information flow vector and the propagation matrices for the elementary gates are needed for the classical information processing, and the total number of steps are taken by a classical computer in parallel for doing this is the total number of gates in a quantum circuit under simulation. Furthermore, no preprocessing and no additional computational steps are required for classical information processing in the HQCM. 

We also have shown that how one can realize efficiently the multi-control gates (like $\Lambda^{12...(n-1)}Z^{n}$) with the HQCM, which play a very important role in the implementation of GA. The gate $\Lambda^{12...(n-1)}Z^{n}$ together with the Hadamard and the $X$ gates is sufficient to realize GA in the framework of HQCM.

One can carry on the investigation in the following directions. In addition to the multi-qubit rotations in the set of elementary gates for the HQCM, one could include some more gates---which can be executed in one shot by the MQCM without adding further complications in the model---in the set of elementary gates. Further, implementation errors need to be considered for any practical realization. In order to establish the fault-tolerance version of HQCM, one needs to design the elementary gates in a fault-tolerant manner but this is beyond the scope of the present paper.

%%%%%%%%%%%%%%%%%%%%%%%%%%%%%%%%%%%%%%%%%%%%%%%%%%%%%%%%%%%%%%%%%%%%%%%%%%%%%%%%%%%%%%%%%%%%%%%%%%%%%%%%%%%%%%%%%%%%%%%%%%%%%%%%%%%%

%%%%%%%%%%%%%%%%%%%%%%%%%%%%%%%%%%%%%%%%%%%                                                %%%%%%%%%%%%%%%%%%%%%%%%%%%%%%%%%%%%%%%%%
%%%%%%%%%%%%%%%%%%%%%%%%%%%%%%%%%%%%%%%%%%%                 Appendix                            %%%%%%%%%%%%%%%%%%%%%%%%%%%%%%%%%%%%%%%%%%
%%%%%%%%%%%%%%%%%%%%%%%%%%%%%%%%%%%%%%%%%%%                                                %%%%%%%%%%%%%%%%%%%%%%%%%%%%%%%%%%%%%%%%%

%%%%%%%%%%%%%%%%%%%%%%%%%%%%%%%%%%%%%%%%%%%%%%%%%%%%%%%%%%%%%%%%%%%%%%%%%%%%%%%%%%%%%%%%%%%%%%%%%%%%%%%%%%%%%%%%%%%%%%%%%%%%%%%%%%%%

\appendix
%%%%%%%%%%%%%%%%%%%%%%%%%%%%%%%%%%%%%%%%%%%                                                %%%%%%%%%%%%%%%%%%%%%%%%%%%%%%%%%%%%%%%%%
%%%%%%%%%%%%%%%%%%%%%%%%%%%%%%%%                  Single-qubit rotation                        %%%%%%%%%%%%%%%%%%%%%%%%%%%%%%%%%%%%
%%%%%%%%%%%%%%%%%%%%%%%%%%%%%%%%%%%%%%%%%%%                                                %%%%%%%%%%%%%%%%%%%%%%%%%%%%%%%%%%%%%%%%%

\section{\label{sec:SQR} Single-qubit rotation $R_{z}(\varphi)$}
\begin{figure}[ht]% h=here, t=top, p=special page
\centering
\includegraphics[trim = 50mm 90mm 50mm 80mm, clip=true , width=0.8\linewidth]{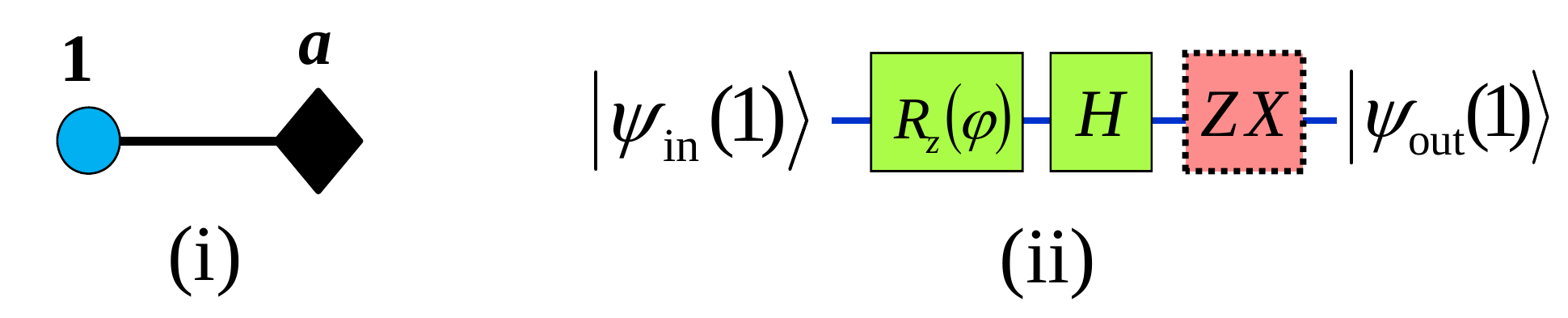}
 %l from the left, b from the bottom, r from the right, and t from the top. Where l, b, r and t are lengths.
\caption{(Color online) (i) The graph associated to the graph state $\arrowvert \phi\rangle_{(1+1)}$ given by Eq.~(\ref{g1}). The (blue) circle, the bond, and the (black) diamond represent the logical qubit which is in the input state $\arrowvert \psi_{\mathrm{in}}(1)\rangle$, the $\textsc{cz}$ operations given by Eq.~(\ref{cz}), and the ancilla qubit $a$, respectively. (ii) The quantum circuit illustrates the effect on the input state, when the qubit `$a$' is measured in an appropriately chosen basis.}\label{fig:5}
\end{figure}

In this appendix, we discuss the implementation of the single-qubit rotation around the $z$-axis $R_{z}(\varphi) = \exp\left(-i\varphi Z/2\right)$ with the MQCM \cite{Raussendorf01, Raussendorf011}. A two-qubit graph state corresponding to the graph depicted in Fig.~\ref{fig:5}(i) is sufficient for accomplishing the job. The logical qubit 1 (represented by the circle) is in a general input state $\left|\psi_{\mathrm{in}}(1)\right\rangle$, and this is the single-qubit state on which we want to apply $R_{z}(\varphi)$. In order to generate the required graph state, we prepare the qubit $a$ (represented by the diamond) in the state $\left(\left|0\right\rangle_{a}+(-1)^{\kappa_{a}}\left|1\right\rangle_{a}\right)/\sqrt{2}$. Then both the qubits are connected by the operation $\textsc{cz}$, which is represented by the bond in Fig.~\ref{fig:5}(i) and given by Eq.~(\ref{cz}). The resulting graph state 
%%%
\begin{equation}
\arrowvert \phi\rangle_{(1+1)}=\frac{1}{\sqrt{2}} \left[\arrowvert 0\rangle_{a}\otimes\arrowvert \psi_{\mathrm{in}}(1)\rangle+(-1)^{\kappa_{a}}\arrowvert 1\rangle_{a}\otimes\left(Z\arrowvert \psi_{\mathrm{in}}(1)\rangle\right)\right] 
\label{g1}
\end{equation}
%%%
is ready for the computation. 
Here, the subscript 1+1 indicates that this graph state is made of two qubits, the logical qubit 1 and the ancilla qubit $a$.

In order to generate the desired effect on the input state, we measure the qubit 1 in the basis 
\begin{equation}
\left\{\arrowvert\uparrow,\downarrow(\pi/2,-\varphi)\rangle_{1}\right\}=\left\{\left(\left|0\right\rangle_{1}+(-1)^{m_{1}}e^{-i\varphi}\left|1\right\rangle_{1}\right)/\sqrt{2}\right\}, \label{basisR}
\end{equation}
and the value of $m_{1}$ is the result of the measurement. After the measurement, the output state (up to a global phase) 
%%%
\begin{equation}
\arrowvert \psi_{\mathrm{out}}(1)\rangle=\left(X \right)^{m_{1}}\left(Z\right)^{\kappa_{a}}H R_{z}(\varphi)\arrowvert \psi_{\mathrm{in}}(1)\rangle 
\label{out1}
\end{equation}
%%%
is obtained from the qubit $a$, and the qubit 1 gets projected either onto the state $\arrowvert \uparrow(\pi/2,-\varphi)\rangle_{1}$ (if $m_{1}=0$) or onto the state $\arrowvert\downarrow(\pi/2,-\varphi)\rangle_{1}$ (if $m_{1}=1$). The net effect on the input state is the required operation $R_{z}(\varphi)$ followed by the Hadamard gate $$H=(X+Z)/\sqrt{2}$$ (represented by the boxes in Fig.~\ref{fig:5}(ii)) and the byproduct operator $\left(X \right)^{m_{1}}\left(Z\right)^{\kappa_{a}}$ (represented by the dotted-box in Fig.~\ref{fig:5}(ii)).

%%%%%%%%%%%%%%%%%%%%%%%%%%%%%%%%%%%%%%%%%%%%%%%       FIGURE (5)            %%%%%%%%%%%%%%%%%%%%%%%%%%%%%%%%%%%%%%%%%%%%%%%%%

%%%%%%%%%%%%%%%%%%%%%%%%%%%%%%%%%%%%%%%%%%%%%%%%%%%%%%%%%%%%%%%%%%%%%%%%%%%%%%%%%%%%%%%%%%%%%%%%%%%%%%%%%%%%%%%%%%%%%%%%%%%%

Here, the direction of measurement lies in $x$,$y$ plane of the Bloch sphere, and the input information is not only teleported from one lattice site to other but also gets processed by the measurement. As we know, every rotation in the Bloch sphere corresponds to a single-qubit operation (up to a global phase). Owing to the Euler decomposition of an arbitrary rotation $R(\alpha,\beta,\gamma)=R_{z}(\gamma)R_{x}(\beta)R_{z}(\alpha)$, where $R_{x}(\beta) = \exp\left(-i\beta X/2\right)$, one can generate an arbitrary single-qubit operation with a chain of five qubits graph state with the measurement direction for each qubit (the angles $\alpha, \beta, \gamma$) in $x$,$y$ plane of the Bloch sphere \cite{Raussendorf01, Raussendorf011}. The realization of an arbitrary rotation by such a sequence of two $z$ rotations sandwiching an $x$ rotation illustrates the importance of the temporal ordering of the measurements in the MQCM.

%%%%%%%%%%%%%%%%%%%%%%%%%%%%%%%%%%%%%%%%%%%                                                %%%%%%%%%%%%%%%%%%%%%%%%%%%%%%%%%%%%%%%%%
%%%%%%%%%%%%%%%%%%%%%%%%%%%%%%%%                         propagation matrices                   %%%%%%%%%%%%%%%%%%%%%%%%%%%%%%%%%%%%
%%%%%%%%%%%%%%%%%%%%%%%%%%%%%%%%%%%%%%%%%%%                                                %%%%%%%%%%%%%%%%%%%%%%%%%%%%%%%%%%%%%%%%%

\section{\label{sec:PM} propagation matrices}
In this appendix, the propagation matrices for the $R$-, $\textsc{cnot}$-, $H$-, and $R_{z}(\pi/2)$-gate for the case of $n$ logical qubits are given. The propagation matrix $\textbf{C}$ is a $2n\times2n$ matrix of the form $\left(
\begin{array}{c|c}
\textbf{C}_{xx} & \textbf{C}_{zx}\\
\hline
\textbf{C}_{xz} & \textbf{C}_{zz}\\
\end{array}
\right)$,
where $\textbf{C}_{xx}, \textbf{C}_{zx}, \textbf{C}_{xz}$ and $\textbf{C}_{zz}$ are $n\times n$ matrices with binary-valued entries \cite{Raussendorf02}. One can generate the propagation matrices for an arbitrary single-qubit rotation $R^{(j)}$ on the logical qubit `$j$' with
\begin{align}
\left[\textbf{C}_{xx}(R^{(j)})\right]_{kl}&= \left[\textbf{C}_{zz}(R^{(j)})\right]_{kl}=\delta_{kl},\nonumber\\
\left[\textbf{C}_{zx}(R^{(j)})\right]_{kl}&= \left[\textbf{C}_{xz}(R^{(j)})\right]_{kl}=0;\label{CCR}
\end{align}
Here, $\left[\textbf{C}_{xx}(R^{(j)})\right]_{kl}$ stands for the entry in $k$th row and $l$th column of the matrix $\textbf{C}_{xx}$ corresponding to the $R^{(j)}$ gate. The same notation apply for the propagation matrices defined below. 

The propagation matrix for the $\textsc{cnot}(a, b)$ gate (both the control qubit `$a$' and the target qubit `$b$' belong to the set of $n$ logical qubits; $a\neq b$) is given by
\begin{align}
\left[\textbf{C}_{xx}(\textsc{cnot}(a, b))\right]_{kl}&= \delta_{kl}+\delta_{kb}\delta_{la},\nonumber\\
\left[\textbf{C}_{zz}(\textsc{cnot}(a, b))\right]_{kl}&=\delta_{kl}+\delta_{ka}\delta_{lb},\nonumber\\
\left[\textbf{C}_{zx}(\textsc{cnot}(a, b))\right]_{kl}&=\left[\textbf{C}_{xz}(\textsc{cnot}(a, b))\right]_{kl}= 0;\label{CCcnot}
\end{align}
for the Hadamard gate $H^{(j)}$ on the logical qubit `$j$' is given by
\begin{align}
\left[\textbf{C}_{xx}(H^{(j)})\right]_{kl}&= \left[\textbf{C}_{zz}(H^{(j)})\right]_{kl}=\delta_{kl}\oplus\delta_{kj}\delta_{lj},\nonumber\\
\left[\textbf{C}_{zx}(H^{(j)})\right]_{kl}&= \left[\textbf{C}_{xz}(H^{(j)})\right]_{kl}=\delta_{kj}\delta_{lj};\label{CCH}
\end{align}
and for the $\pi/2$-phase gate $R^{(j)}_{z}(\pi/2)$ on the logical qubit `$j$' is given by 
\begin{align}
\left[\textbf{C}_{xx}(R^{(j)}_{z}(\pi/2))\right]_{kl}&= \left[\textbf{C}_{zz}(R^{(j)}_{z}(\pi/2))\right]_{kl}=\delta_{kl},\nonumber\\
\left[\textbf{C}_{zx}(R^{(j)}_{z}(\pi/2))\right]_{kl}&= 0,\nonumber\\
\left[\textbf{C}_{xz}(R^{(j)}_{z}(\pi/2))\right]_{kl}&=\delta_{kj}\delta_{lj}.\label{CCphase}
\end{align}
It is advantageous to deal with the information flow vector $\mathcal{I}$ together with the propagation matrices (Eqs.~(\ref{CCR})--(\ref{CCphase})) by a classical computer, than directly dealing with the corresponding byproduct operator $\textbf{\textit{U}}_{B}$ together with the propagation relations (Eqs.~(\ref{proR}), (\ref{proCNOT}), (\ref{proH}), and (\ref{proPhase})).

%%%%%%%%%%%%%%%%%%%%%%%%%%%%%%%%%%%%%%%%%%%                                                %%%%%%%%%%%%%%%%%%%%%%%%%%%%%%%%%%%%%%%%%
%%%%%%%%%%%%%%%%%%%%%%%%%%%%%%%%                   Effective measurement scheme                %%%%%%%%%%%%%%%%%%%%%%%%%%%%%%%%%%%%
%%%%%%%%%%%%%%%%%%%%%%%%%%%%%%%%%%%%%%%%%%%                                                %%%%%%%%%%%%%%%%%%%%%%%%%%%%%%%%%%%%%%%%%

\section{\label{sec:EMS} Efficient measurement scheme}

This appendix holds a discuss on an efficient measurement scheme of the MQCM where the temporal order of the measurements plays an important role \cite{Raussendorf02}. On one hand in the UQCM, we cannot parallelize two gates of a sequence that do not commute. On the other hand in the MQCM, all the gates from the generating set of the Clifford group can be executed in a single time step irrespective of their positions in the circuit. In other words, the temporal order of the measurements in the MQCM is not pre-imposed by the temporal order of the gates. So, the most efficient scheme for the measurements does not necessarily follow the temporal order of the gates in a circuit under simulation. First, the spatial pattern of the measurement bases is assigned to the graph qubits according to the sequence of gates. Then the measurements are performed round by round according to the scheme which is given as follows. 
  
First, the graph $\mathcal{G}$ is divided into disjoint subsets of qubits $\mathcal{Q}_t$, where index $t$ stands for the round of measurements and $0\leq t\leq t_{\mathrm{max}}$. Mathematically, $\mathcal{G}=\bigcup^{t_{\mathrm{max}}}_{t=0}\mathcal{Q}_t$ and $\mathcal{Q}_s \cap\mathcal{Q}_t=\emptyset$ for all $s\neq t$ . The subset $\mathcal{Q}_t$ is a collection of all those qubits which will be measured simultaneously in $t$th round. All the measurements in the $X$, $Y$ and $Z$ eigenbasis are put together in the very first round ($0$th round), and there is no need for adjusting the measurement bases according to the previous measurement results for the qubits of $\mathcal{Q}_0$. In the first measurement round, the ``redundant graph qubits'' are removed by the $Z$-measurements, the ``readout qubits'' are measured in the $Z$ eigenbasis, and the Clifford part of the circuit is executed by the $X$-, $Y$-measurements \cite{Raussendorf02}. In the MQCM, the ``readout qubits'' which play the role of ``output register'' are not the last ones to be measured, they are among the first ones. 

The observables for all subsequent measurement rounds are of the form $\cos(\varphi)X \pm \sin(\varphi)Y$ with $\varphi\notin\left\{0, \pm\frac{\pi}{2}\right\}$, where the measurement bases for the qubits are decided by the measurement outcomes from the previous rounds. All those qubits whose measurement bases depend on the outcomes from the first measurement round belong to the next subset $\mathcal{Q}_1$. Similarly, the measurement outcomes from the subset $\mathcal{Q}_1$ together with $\mathcal{Q}_0$ decide the measurement bases for the qubits in $\mathcal{Q}_2$, and so on. These subsets are measured one by one up to the final measurement round $t_{\mathrm{max}}$. One can think of the total number of measurement rounds ($t_{\mathrm{max}}+1$) as the logical depth (temporal complexity) for the MQCM. 

Parallel to the measurement rounds, the classical data-processing parts are taken care of by a classical computer. After preparing the graph state and just before starting the measurements, the information vector is initialized to $\mathcal{I}^{\textsc{mqcm}}_{\mathrm{init}}$. $\mathcal{I}^{\textsc{mqcm}}_{\mathrm{init}}$ depends on the eigenvalues $\left\{\kappa\right\}$ of the graph and some particular gates (like $\textsc{cnot}$, $R_{z}(\pi/2)$) which appear in a quantum circuit under simulation \cite{Raussendorf02}. After executing the first measurement round on the set $\mathcal{Q}_0$, $\mathcal{I}^{\textsc{mqcm}}_{\mathrm{init}}$ gets updated to $\mathcal{I}(0)$ through the measurement results. $\mathcal{I}(0)$ then determines the measurement bases for the qubits of $\mathcal{Q}_1$. Similarly, the measurement outcomes from round $t$ update the information flow vector from $\mathcal{I}(t-1)$ to $\mathcal{I}(t)$. The corresponding byproduct operator  is given by 
\begin{equation}
\textbf{\textit{U}}_{B}(t)=\prod^{n}_{j=1}(X^{(j)})^{x_{j}(t)}(Z^{(j)})^{z_{j}(t)}.
\label{byproMBQCM} 
\end{equation} 
Then, $\mathcal{I}(t)=\mathcal{I}({x_{j}(t), z_{j}(t)})$ sets the measurement bases for the $(t+1)$th round. After the final measurement round $t_{\mathrm{max}}$, the $x$-part of the information flow vector $\mathcal{I}_{x}(t_{\mathrm{max}})$ enables us to interpret the result of the computation.

In this measurement scheme, the following technical points are worth emphasizing, which are discussed in Ref.~\cite{Raussendorf02}: (1) In order to construct the subsets of graph qubits $\left\{\mathcal{Q}_t\right\}$, a classical computer needs the forward cones for all the graph qubits. The forward cones decide a strict partial ordering among the qubits, and the sets $\left\{\mathcal{Q}_t\right\}$ are constructed accordingly. (2) In order to account for the influence of the measurement outcomes and the set of eigenvalues $\left\{\kappa\right\}$ on $\mathcal{I}(t)$, a classical computer needs the byproduct images for all the graph qubits. (3) $\left\{\kappa\right\}$, the byproduct images, and $\mathcal{I}(t)$ are required for setting the measurement bases for the as-yet unmeasured qubits. A classical computer uses the symplectic scalar product for doing this.

%%%%%%%%%%%%%%%%%%%%%%%%%%%%%%%%%%%%%%%%%%%%%%%%%%%%%%%%%%%%%%%%%%%%%%%%%%%%%%%%%%%%%%%%%%%%%%%%%%%%%%%%%%%%%%%%%%%%%%

%%%%%%%%%%%%%%%%%%%%%%%%%%%%%%%%%%%%%%%              Acknowledgements           %%%%%%%%%%%%%%%%%%%%%%%%%%%%%%%%%%%%%%%%%%

%%%%%%%%%%%%%%%%%%%%%%%%%%%%%%%%%%%%%%%%%%%%%%%%%%%%%%%%%%%%%%%%%%%%%%%%%%%%%%%%%%%%%%%%%%%%%%%%%%%%%%%%%%%%%%%%%%%%%%
\begin{acknowledgements}
Centre for Quantum Technologies is a Research Centre of Excellence funded by Ministry of Education and National Research Foundation of Singapore. D. Z. would like to express his gratitude to CQT for its kind hospitality, and further thanks to H. J. Briegel and W. D\"{u}r. A. S. wishes to thank H. J. Briegel and IQOQI for their warm hospitality. A. S. also wishes to express his grateful feelings to W. D\"{u}r, Hui Khoon, Le Huy Nguyen, and Philippe Raynal for enlightening comments and friendly discussions.
\end{acknowledgements}
%%%%%%%%%%%%%%%%%%%%%%%%%%%%%%%%%%%%%%%%%%%%%%%%%%%%%%%%%%%%%%%%%%%%%%%%%%%%%%%%%%%%%%%%%%%%%%%%%%%%%%%%%%%%%%%%%%%%%%

%%%%%%%%%%%%%%%%%%%%%%%%%%%%%%%%%%%%%%%              Bibliography             %%%%%%%%%%%%%%%%%%%%%%%%%%%%%%%%%%%%%%%%%%

%%%%%%%%%%%%%%%%%%%%%%%%%%%%%%%%%%%%%%%%%%%%%%%%%%%%%%%%%%%%%%%%%%%%%%%%%%%%%%%%%%%%%%%%%%%%%%%%%%%%%%%%%%%%%%%%%%%%%%

\end{document}